\documentclass[aps,prl,amssymb,superscriptaddress]{revtex4}
\usepackage{graphicx}
\usepackage{dcolumn}
\usepackage{bm}
\usepackage[dvips]{epsfig}

\begin{document}

\title{Imaginary crystals made real}

\date{\today}

\author{Simone Taioli}
\affiliation{Faculty of Mathematics and Physics, Charles University in Prague, Czech Republic}
\affiliation{European Centre for Theoretical Studies in Nuclear Physics and Related Areas (ECT*), Bruno Kessler Foundation \& Trento Institute for Fundamental Physics and Applications (TIFPA-INFN), Trento, Italy}
\author{Ruggero Gabbrielli}
\affiliation{Faculty of Mathematics and Physics, Charles University in Prague, Czech Republic}
\author{Stefano Simonucci}
\affiliation{Department of Physics, University of Camerino, Italy \& Istituto Nazionale di Fisica Nucleare, Sezione di Perugia, Italy}
\author{Nicola Maria Pugno}
\affiliation{Laboratory of Bio-inspired \& Graphene Nanomechanics, Department of Civil, Environmental and Mechanical Engineering, University of Trento, Italy}
\affiliation{School of Engineering and Materials Science, Queen Mary University of London, UK}
\affiliation{Center for Materials and Microsystems, Bruno Kessler Foundation, Trento, Italy}
\author{Alfredo Iorio}
\affiliation{Faculty of Mathematics and Physics, Charles University in Prague, Czech Republic}


\begin{abstract}
We realize Lobachevsky geometry in a simulation lab, by producing a carbon-based mechanically stable molecular structure, arranged in the shape of a Beltrami pseudosphere. We find that this structure: i) corresponds to a non-Euclidean crystallographic group, namely a loxodromic subgroup of $SL(2,\mathbb{Z})$; ii) has an unavoidable singular boundary, that we fully take into account. Our approach, substantiated by extensive numerical simulations of  Beltrami pseudospheres of different size, might be applied to other surfaces of constant negative Gaussian curvature, and points to a general procedure to generate them. Our results also pave the way to test certain scenarios of the physics of curved spacetimes.
\end{abstract}

\maketitle

Lobachevsky used to call his Non-Euclidean geometry ``imaginary geometry'' \cite{kaufmann1981}. Beltrami showed that this geometry can be realized in our Euclidean 3-space, through surfaces of \textit{constant negative} Gaussian curvature $K$ \cite{beltrami}. Thus, next to the one smooth surface of constant positive $K$, the sphere, we had to add infinitely many \textit{singular} surfaces of constant negative $K$. The unavoidable singularities descend from the Hilbert theorem, stating that no analytic complete (smooth) surface of constant negative $K$ can exist in the Euclidean 3-space \cite{ovchinnikov}.

Just as the discrete subgroups of the symmetries of the sphere, $SO(3)$, are in close connection with the symmetries of crystals and molecules \cite{hamermesh}, the discrete subgroups of the symmetries of the Lobachevsky plane, $SO(2,1) \sim SL(2,\mathbb{R})$, are the Non-Euclidean crystallographic (NEC) groups \cite{NECgroups}. Nonetheless, the physical significance of the latter, in real molecular or lattice structures, is obscure \footnote{Experimental triangulations of colloidal surfaces of negative curvature exist \cite{irvine}. However, they are not focused on the Lobachevsky plane (constant $K$). Theoretical triangulations of the Lobachevsky plane also exist \cite{nelsonnegative}, but the embedding into $\mathbb{R}^3$ is not considered.}. Our aim here is to explore their actual realization, by facing the non-trivial effects of the Hilbert theorem, and by focusing on a (carbon-made, graphene) Beltrami pseudosphere, that realizes portions of the Lobachevsky plane in our Euclidean 3-space, while keeping some of the symmetries of the sphere (see Supplemental Material [SM]).
\begin{figure}[h]
\includegraphics[height=8cm, trim={0 0.8 0. 0.8},clip,angle=-90]{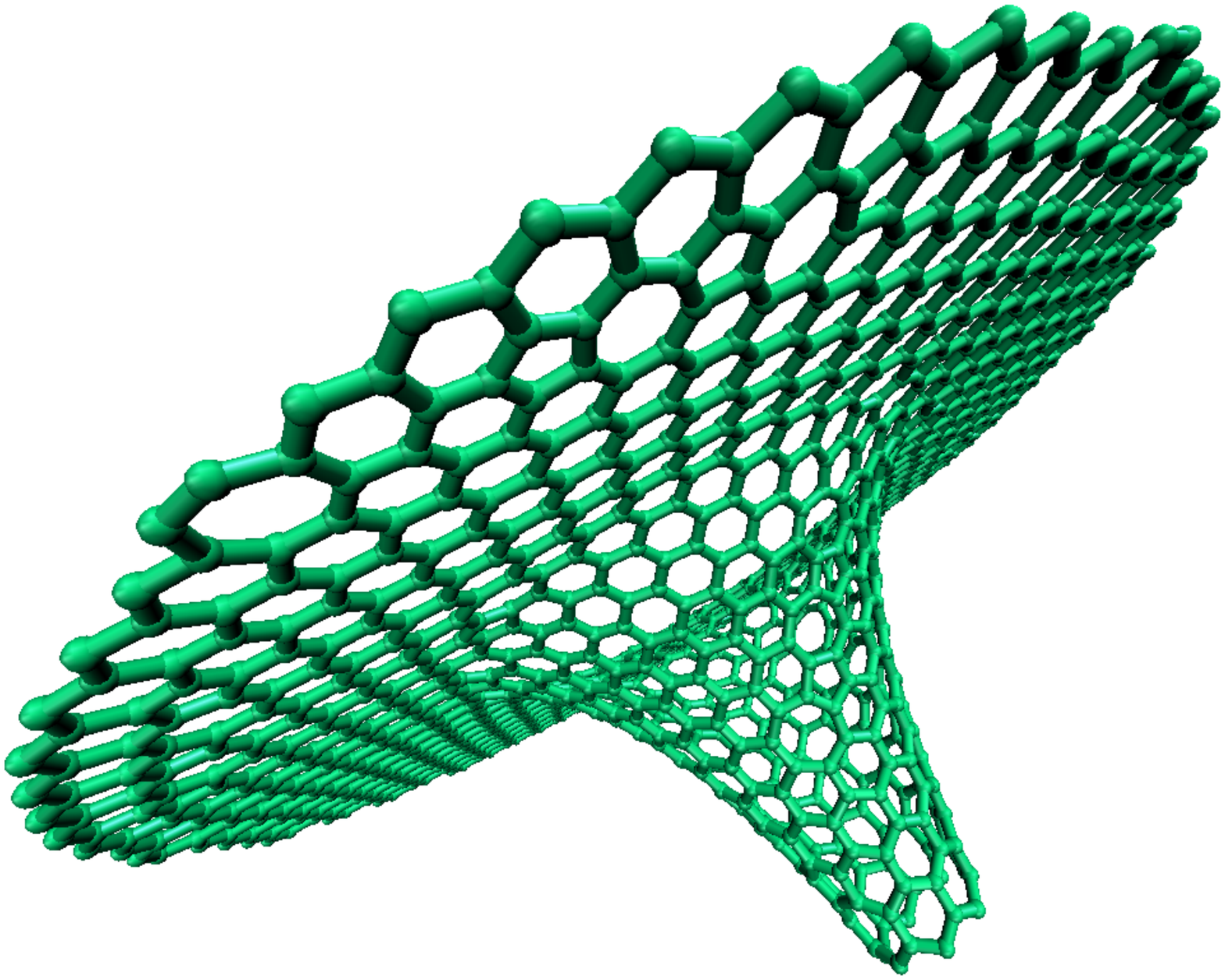}
 \caption{\label{trumpet1} Ideal tiling of a truncated Beltrami pseudosphere, realized with a trivalent lattice.}
 \end{figure}
Besides the mathematical charm, and the focus on an important material, our study is beneficial for many research areas, ranging from material science to biology~\cite{bio}, and from the discrete structures of curved spacetimes \cite{mauro} to the generalized Thomson problem \cite{Thomson}. The choice of graphene is also motivated by the recently proposed occurrence of a Hawking effect on a carbon-made Beltrami pseudosphere \cite{iorio} (see \cite{iorioreview} for a review).

Let us assume that only trivalent lattices are allowed. (In general, this is not strictly necessary but, besides being natural for graphene, this assumption simplifies the discussion). With this, to tile the sphere ($K = r^{-2}$) we need either triangles, or squares or pentagons; to tile the flat plane ($K = 0$) we need only the hexagon; to tile the Lobachevsky plane ($K = - r^{-2}$), whose line element is
\begin{equation}
d l^2 = \frac{r^2}{{\tilde y}^2}(d{\tilde x}^2+d{\tilde y}^2) \;, \quad \tilde{y} > 0 \;, \label{lobsurface}
\end{equation}
we need one of the infinitely many other polygons: the heptagon, the octagon, etc. This descends from the Euler-Poincar\`e formula and the Gauss-Bonnet theorem for manifolds $\Sigma$ without boundaries \cite{Eisenhart}
\begin{equation}
K_{tot} = 2 \pi \chi  = \phi \sum_{n_s} n_s (6 - s) \label{GBandEuler} \;,
\end{equation}
where $K_{tot} = \int_\Sigma K d^2\mu $ is the total Gaussian curvature of $\Sigma$, $\chi$ its Euler characteristic, $\phi = \pi/3$ when $\Sigma$ is embedded in $\mathbb{R}^3$, $n_s$ is the number of $s$-sided polygons, necessary to tile $\Sigma$. Hence, each $s$-sided polygon, carries a curvature given by $K_s = (6-s) \phi$. This proves our statements.

\begin{figure}[h]
\centering
\includegraphics[width=.9\linewidth]{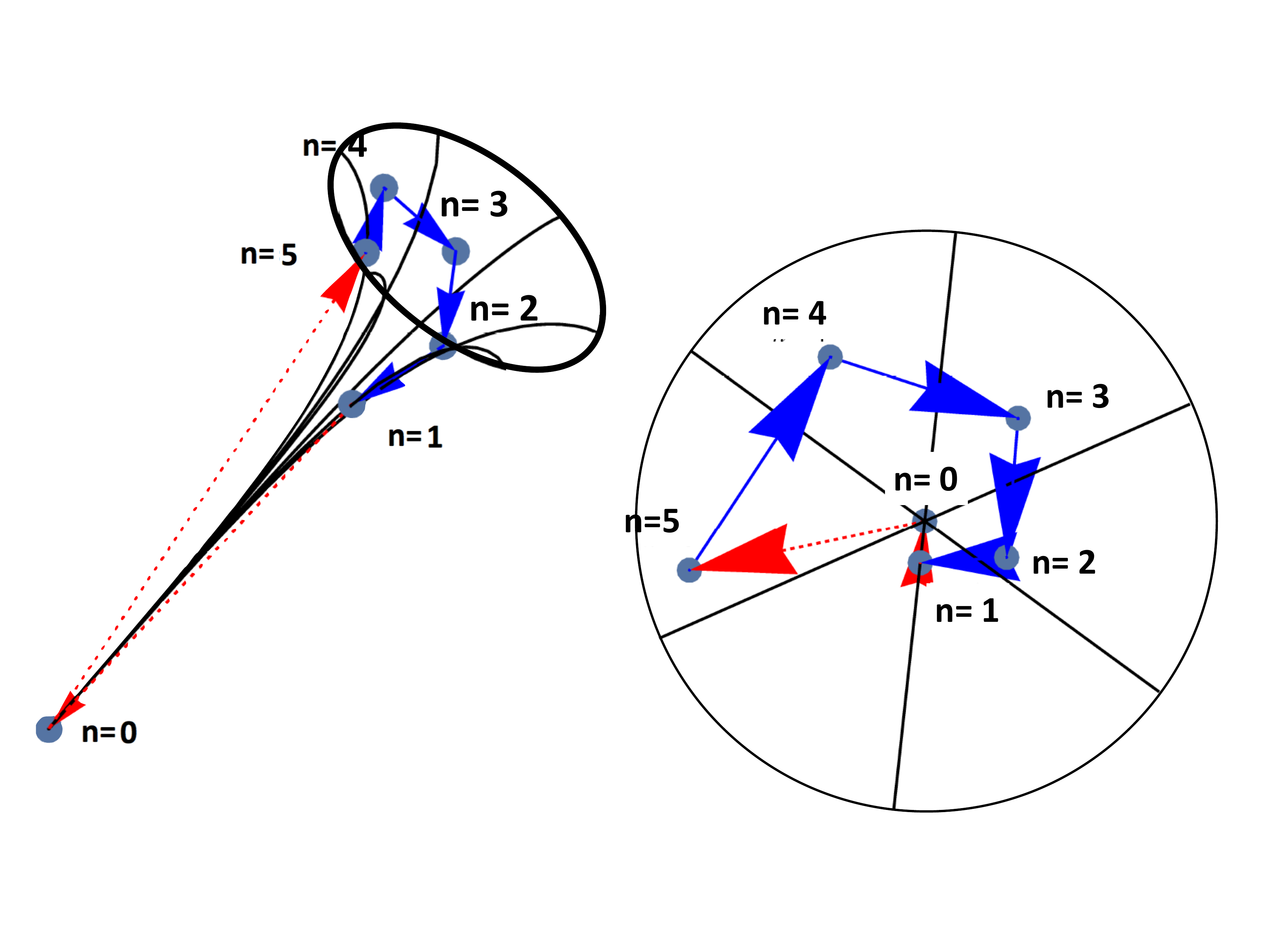}
\caption{\label{loxodromicfuchs} Action of $F_Y$ on the vertices of the ``Lobachevsky Beltrami polyhedron''. The vertex at infinity ($n=0$) is identified with the first vertex ($n=5$). Those vertices are the expected locations of the (excess of) heptagonal defects on the Beltrami pseudosphere. Left: side view; right: top view. }
\end{figure}

To obtain a Beltrami pseudosphere one needs to write the Lobachevsky coordinates in (\ref{lobsurface}) as specific functions of the meridian ($u$) and parallel ($v$) coordinates (with a specific parametrization, see SM) $\tilde{x} = v / r$, and $\tilde{y} = e^{- u / r} / r$, leading to the line element $dl^2 = du^2 + R^2 (u) dv^2$, with $v \in [0, 2\pi]$, and
\begin{equation}\label{explicitxybeltrami}
R(u) = r e^{u/r} \;, \quad u \in [- \infty, 0] \;.
\end{equation}
Clearly, this surface of revolution keeps one rotation symmetry of the sphere ($v \in [0, 2\pi]$), while turning the second rotation into a non-compact one (compare $R(u)$ in (\ref{explicitxybeltrami}) with the sphere's $R(u) = r \cos(u/r)$, $u \in [- r\pi/2, + r \pi/2 ]$).

According to Hilbert theorem, this surface has a singular boundary at $u=0$, that is the maximal circle of radius $R=r$. We call it the \textit{Hilbert horizon} \cite{iorioreview}. This results in a ``decoupling'' of the Gauss-Bonnet theorem and the Euler-Poincar\`e formula, i.e., $K_{tot} = \frac{\pi}{3} \sum_n n_s (6 - s) \neq 2 \pi \chi$, whereas $2 \pi \chi = K_{tot} + \int_{\partial \Sigma} K_g dl$, with $K_g$ the geodesic curvature of the boundary.

We now apply a regular triangular tiling to the surface, and then we shall move to the Voronoi dual. The triangulation is associated with a dense uniform packing, having a fixed lattice spacing $\ell$, and we truncate the surface at a radius of few $\ell$s. Thus, the larger the radius, the finer the discrete approximation to the continuous surface. This proper triangulation is convenient to discuss the theoretical issues as well as to describe the making of the actual carbon pseudosphere shown in Fig.~\ref{trumpet1}. By the previous arguments, we expect for the ideal surface, whose area is $2 \pi r^2$, the number of heptagonal defects to be 6 ($K_{tot} = - 2 \pi = 6 (- \pi / 3)$).

The issue now is to find where these defects are positioned on the surface, and whether the combination of the discrete symmetries of the sphere and of the Lobachevsky plane can guide us to find the symmetries among the defects. In other words, our goal is to construct the negative curvature counterpart of the icosahedral group, $Y = (C_2, C_3, C_5)$, in the case of the Beltrami, say it $F_Y$. $F_Y$ would be one of the infinitely many negative curvature counterpart of $Y$. It would produce a ``Lobachevsky Beltrami polyhedron'', whose vertices are the heptagonal defects (see Fig.~\ref{loxodromicfuchs} and SM).
We apply to $Y$ the same deformations that turn a sphere into a Beltrami, illustrated in details in the SM. The $C_5$ symmetry in the $v$ direction is preserved by construction, while the $C_5$ symmetry in the $u$ direction is ``hyperbolized'' (see (\ref{explicitxybeltrami})), resulting in a discrete ``boost''. Therefore, we need to orient the icosahedron to have one vertex at the south pole of the sphere ($u_{Sphere} = - r \pi / 2$), mapped to the very bottom tip of the Beltrami ($u_{Beltrami} = - \infty$).

The combination of these symmetries provides the following structure of the group
\begin{equation}
F_Y \equiv \{ \exp\{u + i v \} | v = 2 \pi n / 5,  u = r \ln(|n|/6) \} \label{NECgroup}
\end{equation}
where $n \in \mathbb{Z}_5$. The sign of $n$ is related to the spiral's chirality. Structures with opposite chirality have the same energy (spontaneous breaking of the parity transformation of $Y$), as seen in our simulations. As described in detail in the SM, the 6 discrete values of $u = r \ln(R/r)$ correspond to the point at infinity, $R/r = 0$, and to the 5 tangent cones with apertures $\alpha_n = \arcsin (n/6)$, where $n \in \mathbb{Z}_5 - \{ 0 \}$, giving $R/r = 1/6, ..., 5/6$ (in the following figures of the pseudospheres' top view, these radii correspond to colored circles). The point at infinity needs to be removed in all practical realizations, thus $u|_{n =  \pm 5} \equiv u|_{n = 0} = - \infty$.

$F_Y$ is a cyclic loxodromic subgroup of $SL(2, \mathbb{Z})$ of order 5, hence the NEC group of the Beltrami pseudosphere we were looking for.
\begin{figure}[hbt!]
\centering
\includegraphics[width=.6\linewidth]{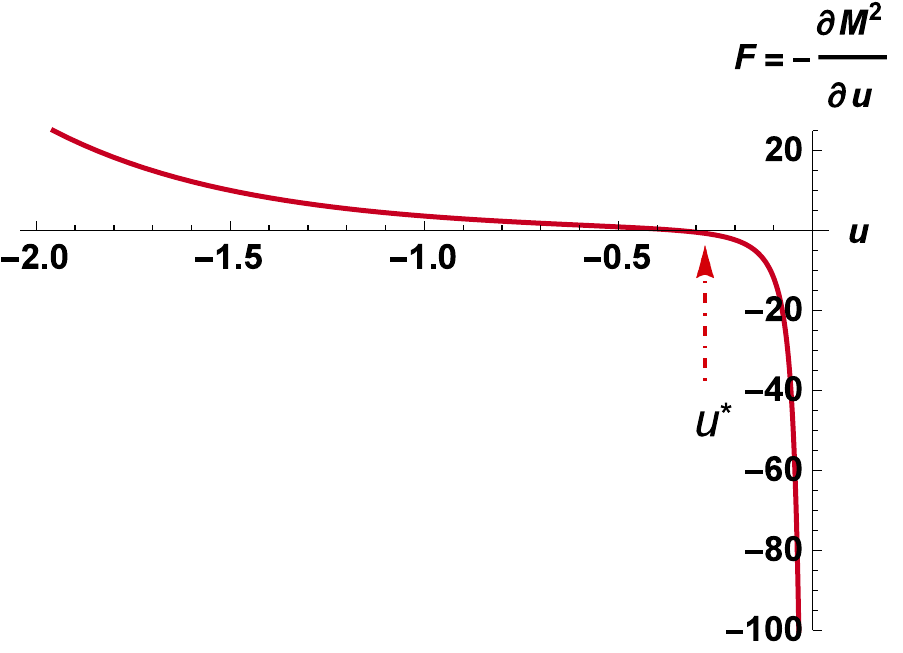}
\caption{Behavior of $F \propto - \partial_u M^2(u)$. The proportionality constant and $r$ are set to 1. At $R(u^*) = r/\sqrt{2}$, $|K_1| = |K_2|$, hence $M = 0 = F$.}
\label{extrinsicforce}
\end{figure}
Real membranes are strained by the force due to their extrinsic curvature, $M = (1/2) (K_1 + K_2)$, that is $\vec{F} \propto - \vec{\nabla} M^2$. For Beltrami the principal curvatures are $K_1 = - \frac{R(u)}{r \sqrt{r^2 - R^2(u)}}$ and $K_2=\frac{\sqrt{r^2 - R^2(u)}}{ r R(u)}$, which give a diverging stretching at the Hilbert horizon, and a diverging contraction at the tip, see Fig.~\ref{extrinsicforce}.
The first divergence is cured by pinning the atoms there with a positive (compressing) force. The second, less severe, divergence is of no concern as the surface is truncated.  This is important to understand the expected formation of ``scars'', that are chains of disclination defects either carrying an overall unit of charge of curvature (positive or negative) or none. Scars come about to relieve the elastic strain of membranes, see, e.g., \cite{Thomson}. Summarizing, in the limit of large numbers of atoms the structures approach the ideal case, and we expect the truncated carbon-made Beltrami pseudospheres to show an excess of 5 heptagonal defects, at the locations predicted by $F_Y$,  on which extrinsic curvature effects act, creating scars along the geodesics, and pushed towards the tip. We shall now describe the robust numerical simulations that showed the correctness of these considerations.

To generate Beltrami pseudosphere equilibrium configurations a number of steps (summarized in Fig. S6 of the SM), were performed.
Surfaces with different $r$s were engineered, aiming at investigating both number and distribution of defects, and at finding the most energetically favourable morphologies. The first step consists in fixing $R_{max}=r$, as well as $R_{min}$, the radius of the smallest circle (truncation), (see Fig. \ref{trumpet1}). Starting from uniformly random distributed $x$ and $y$ planar coordinates, we generate structures initially characterized by a strongly irregular point mesh on the pseudosphere. The number of points $N$ used in the surface tiling of a specific pseudosphere is determined according to the following formula:
\begin{eqnarray}\label{npoints}
 N = 4\pi R_{max} (R_{max}-R_{min})/(\sqrt{3}~a_{CC}^2) \nonumber  + \pi/2 \times \\
\left[ \!1/\arcsin(a_{CC}/(2R_{max}))\!\!+\!1/\arcsin(a_{CC}/(2R_{min}))\right]
\end{eqnarray}
where $a_{CC}=\ell$ is the carbon-to-carbon distance. Notice that $R_{min}$ is such that the distance between opposite carbon atoms in the circumference is larger than $a_{CC}$. To avoid numerical instabilities in the calculation of the repulsive part of the interacting potential, the number of points $N$ should not exceed that required to uniformly cover the entire surface. As such, a predictor-corrector approach, to determine the correct $N$, must be used. Points are constrained, on the one hand, by construction, on the Beltrami surface, and on the other hand, by using steep parabolic potential rumps at both $R_{max}$ and $R_{min}$ (see Fig. \ref{extrinsicforce}). To ensure that the pseudospheres are smoothly terminated (i.e., with zero derivative) at the Hilbert horizon, a planar ring of several $a_{CC}$ in transverse direction is added in the simulations. Thus, potential rumps up steeply only after the flat graphene ring. In general, we find that the bigger is $r$ the larger the flat ring must be. Points on the Beltrami surface interact through a pair-wise Lennard-Jones (LJ) potential, suitable to generate a triangular tiling of the structure (details on LJ are provided in the SM).
 \begin{figure}[hbt!]
\centering
\includegraphics[width=0.5\textwidth]{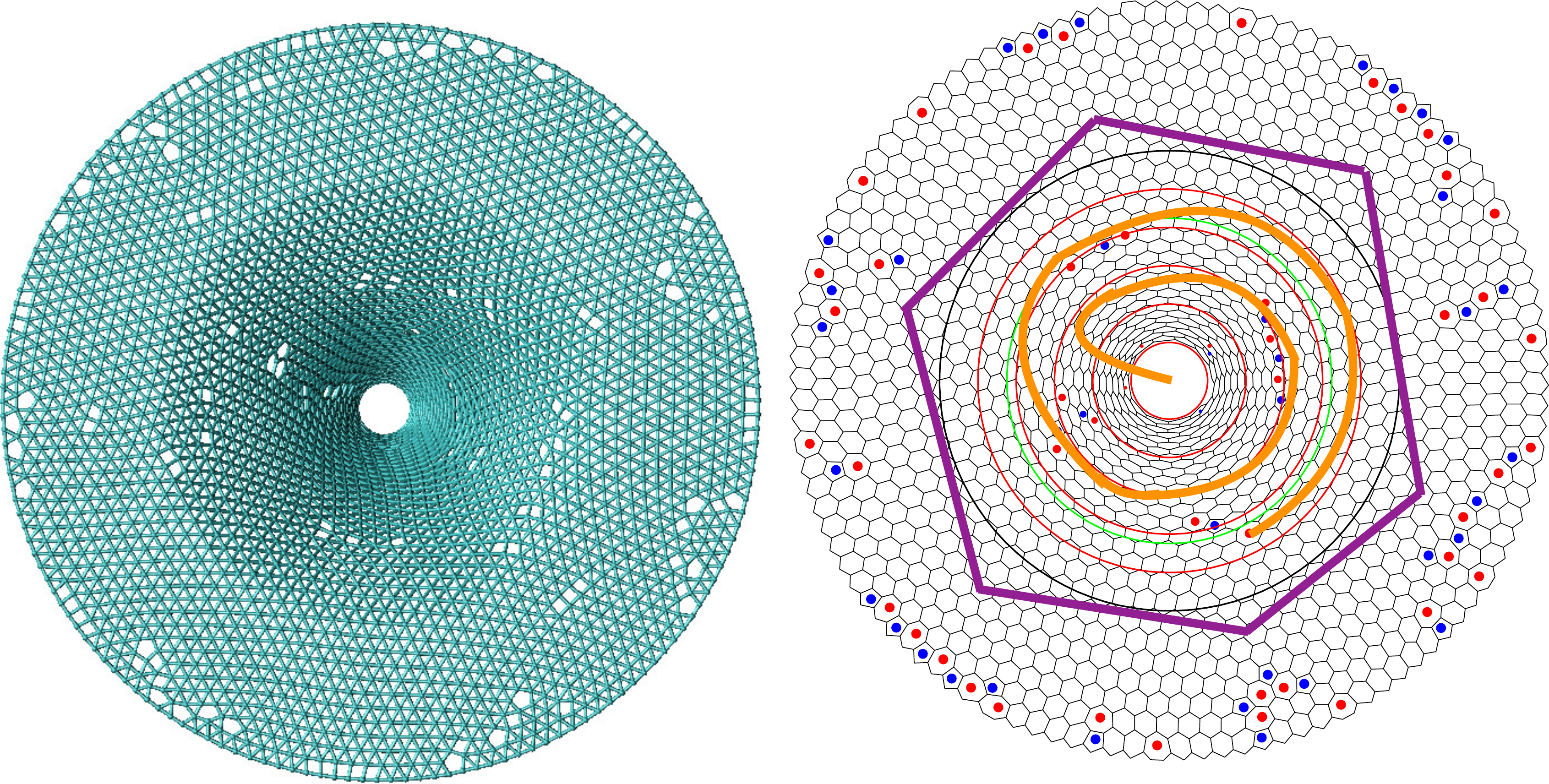}
\caption{\label{tessellation} Left: lattice triangular tessellation obtained by using the FIRE approach along with the LJ interatomic potential; right: scar helicoidal distribution.}
\end{figure}
To find the equilibrium structure of a solid, a variety of well-established approaches are available, such as
quasi-Newton methods (e.g. Broyden-Fletcher-Goldfarb-Shanno algorithm (L-BFGS)), conjugate-gradient (CG) and steepest descent.
However, as previously stated, to mimic the ideal Beltrami structures one needs to decrease the curvature and this results in increasing the size, and thus the number of atoms.
In this case, owing to memory and computational overload, the only viable option to find the structures with minimum potential energy is to rely on the use of the {\it Fast Inertial Relaxation Engine} (FIRE) approach \cite{FIRE} (details on FIRE algorithms and parameters are provided in the SM).
In left panel of Fig. \ref{tessellation} we produce an example of triangulation pattern reachable by FIRE using 2840 points on a Beltrami surface with $R_{max}=36 {\text{~\AA}}$ and $z_{min}=-64 {\text{~\AA}}$, where $z=0$ is the $z$-coordinate of the Hilbert horizon. A number of defects, distributed all over the structure, appear after the LJ triangulation with an excess at the final border of the planar ring. Indeed, the insurgence of defects is more likely whereby the triangular pattern is spoiled.\\
To generate truncated pseudospheres that provide models for $sp^2$-bonded carbon atoms one needs to apply a topological dualization (Voronoi patterning) to the LJ optimized lattice. We initially compute the adjacency matrix of each particle, where a neighbour was defined as a particle closer than $\sqrt{3} \times a_{CC}$ . Distances were evaluated in 3D space and not on the surface. The mesh was then refined in order to output a triangulation. The centres of each triangle were finally exported to a final structure containing pentagonal, hexagonal and heptagonal rings only (see Fig. \ref{trumpet1}).
Heptagonal and pentagonal defects appear to form charged and uncharged linear defects in the form of scars, which tend to distribute over the surface according to a specific helicoidal pattern (see right panel of Fig. \ref{tessellation}). No vertex of valence higher than 3 has been observed when computing charge defect.
In particular, our goal is to discretize the pseudosphere by introducing real carbon atoms. We note that after the Voronoi patterning of the lattice, the carbon structure is far from equilibrium configuration and, thus, we need to optimize the atomic positions on the surface.
However, a LJ model of the interatomic interactions (see Eq. S28 of the SM) cannot reflect the strongly directional $sp^2$ bond of the dualized carbon pseudosphere, which is very much similar to a defective graphene lattice. To cure this LJ pathology, we included three-body angular terms in the interaction potential functional form, using a Stillinger-Weber-type (SW) potential \cite{stillinger}, which has been successfully adopted
in molecular dynamics simulations of graphitic structures \cite{abraham} (see SM for details on SW parameters).
Pseudospheres with a final number of 1146, 2146, and 5506 carbon atoms were generated and optimized, see Fig. \ref{trumpio}.
Some defect topologies were found locally resonating between different configurations (see Fig. S7 of the SM) carrying no net charge (uncharged dipole) as in the case of Stone-Wales defect in planar graphene.
\begin{figure}[hbt!]
\centerline{\includegraphics[width=3cm, angle=90]{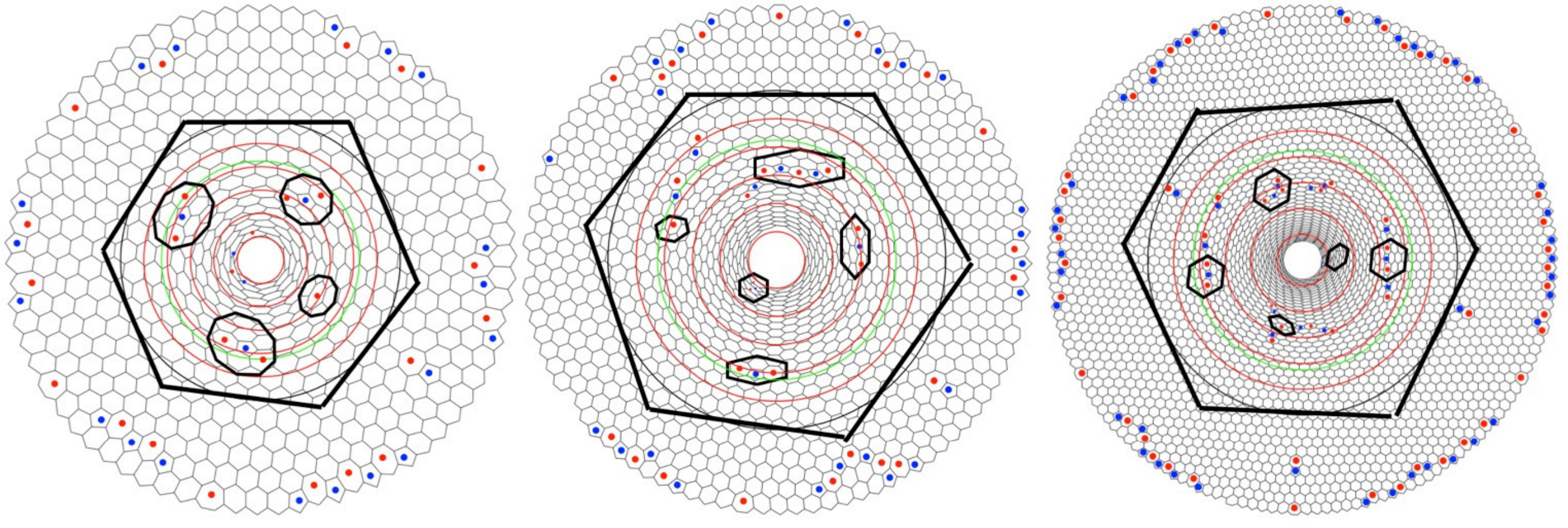}}
\caption{\label{trumpio} Optimized pseudospheres with 1146 (left), 2146 (middle), and 5506 (right) carbon atoms.}
\end{figure}
In this figure, we framed the charged scars (one heptagon in excess). The number of such scars increases with the size of the pseudosphere, converging to the expected five, distributed along a helicoidal path, see Fig. \ref{trumpio}. This is a robust proof to our conjecture on the $F_Y$ symmetry of Beltrami, see Fig. \ref{loxodromicfuchs}. These features  originates from the interplay between local curvature and average inter-particle distance \cite{irvine} in the curved space. Indeed, while it is known that mild curvature leads to the formation of scars, previous studies of the arrangement of particles on the sphere (in the order of 10,000) revealed the formation of icosahedral disclination lines and ground states characterized by high symmetry \cite{Thomson}. Colloidal crystals on negatively curved interfaces such as those of capillary bridges have been observed to form isolated heptagonal defects but no particular symmetry \cite{irvine}. More recent numerical investigations that used a repulsive LJ potential and the basin-hopping method on a number of surfaces with either zero (catenoids) or constant mean curvature (unduloids), report configurations displaying instead some form of symmetry.

\begin{figure}[h]
\includegraphics[width=0.6\linewidth, angle=-90]{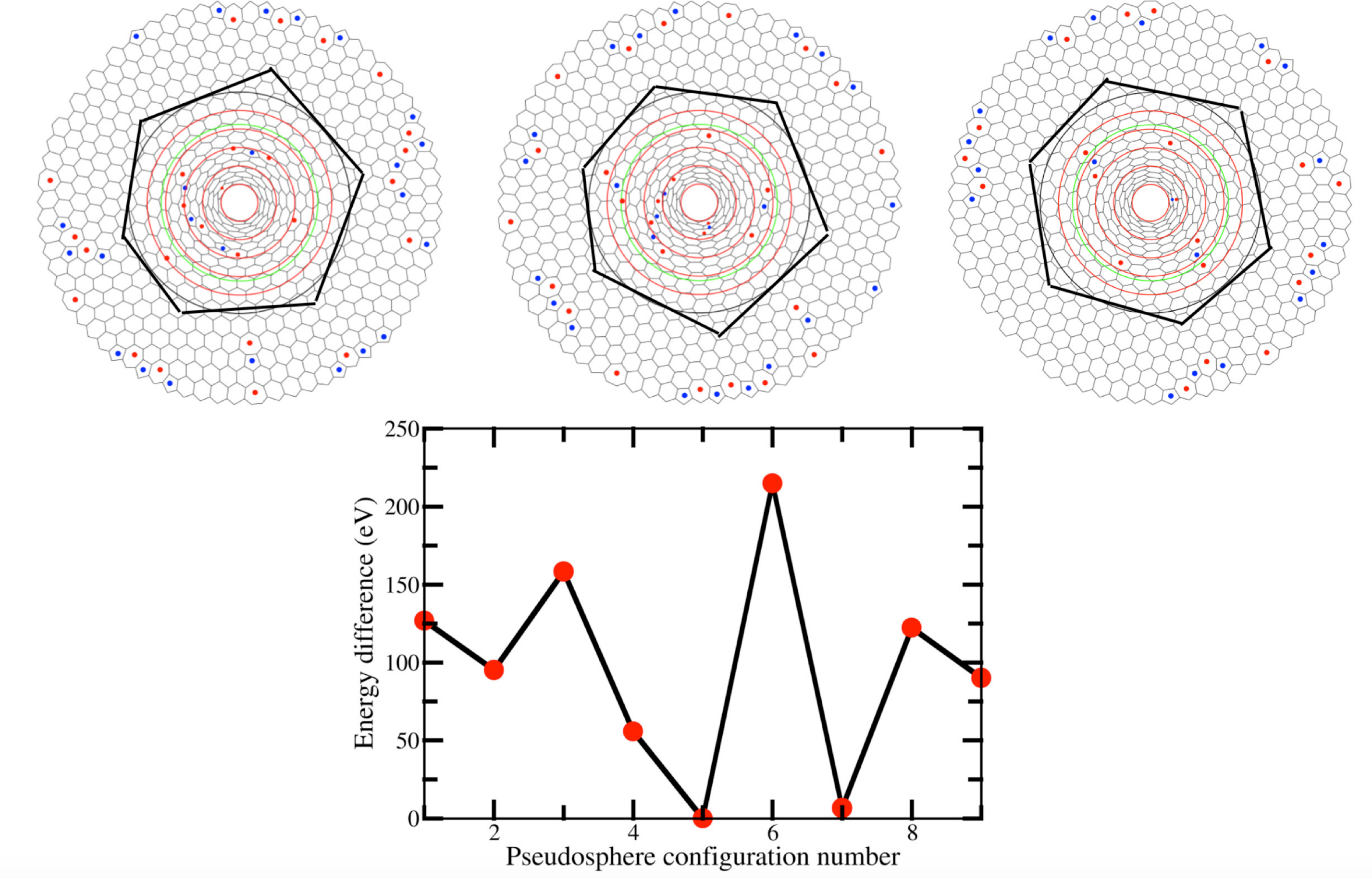}
\caption{\label{1146} Bottom panel: energy landscape for the different Beltrami pseudosphere configurations. Top panel: 3 Beltrami pseudospheres with 1146 atoms obtained by dualization of a 614 point lattice, representing the best MEC candidates. $R_{max}$=16 and $Z_{min}$=-23. }
\end{figure}

\begin{figure}[h]
\includegraphics[width=0.35\linewidth, angle=-90]{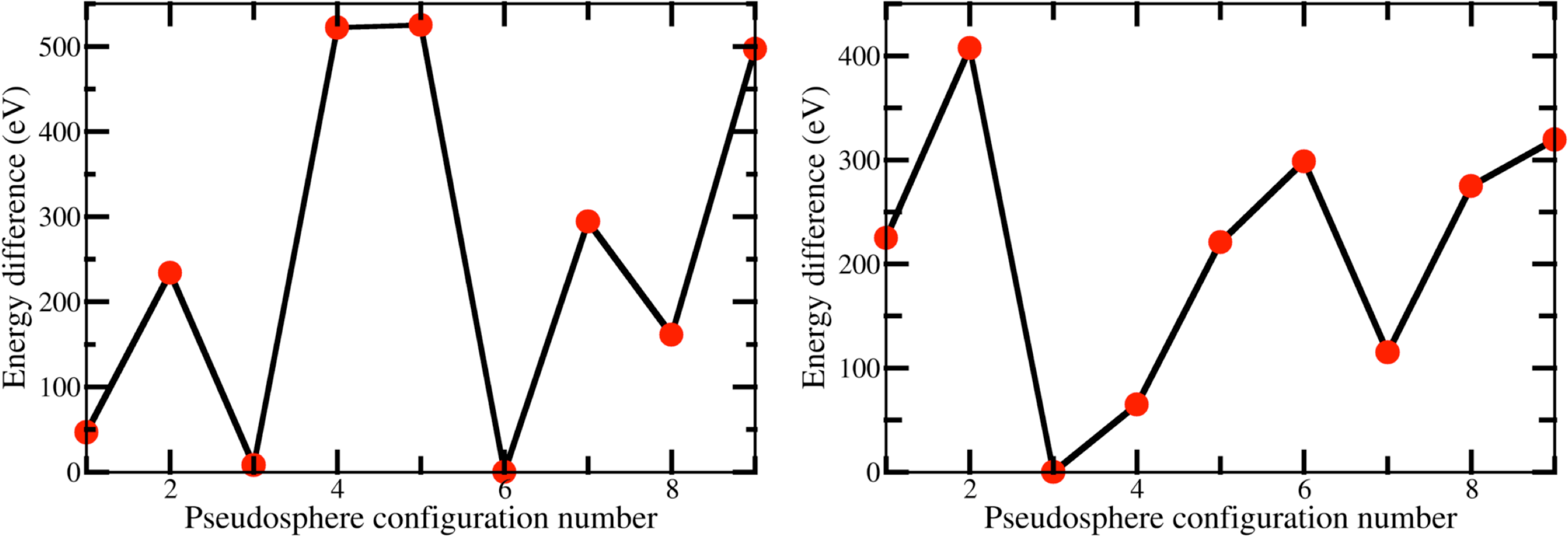}
\caption{\label{2146en} Left: energy landscape for 9 Beltrami pseudospheres with 2146 carbon atoms obtained by dualization of a 1128 point lattice with $R_{max}$=26 and $Z_{min}$=-38. Right: energy landscape for 9 Beltrami pseudospheres with 5506 carbon atoms obtained by dualization of a 2840 point lattice with $R_{max}$=36 and
$Z_{min}$=-64.}
\end{figure}

In Fig. \ref{trumpio} we sketch a large hexagon, whose sides are tangent to the pseudosphere's Hilbert horizon (black circle). The appearance of this hexagonal circulation guarantees that a Beltrami pseudosphere has been generated. This feature has been found only in connection with the absence of charged scars crossing the Hilbert horizon.
Finally, in the presence of crossing events the flat ring surrounding the pseudosphere would bend and deform the Beltrami pseudosphere into a polygonal prismatoid (e.g. for pentagonal circulation we would obtain a pentagonal base pyramid).
In this respect, the inclusion of a graphene flat ring is twofold: it allows pseudosphere generation with zero-derivative singularities, and it removes the interaction of the pseudosphere with the boundaries, thus eliminating the possibility for the latter to interfere with the internal formation of defects (the same effect could be reached by imposing periodic boundary conditions to the pseudosphere). Thus, provided the requirements mentioned above are rigorously satisfied, one can generate properly-defined Beltrami pseudospheres.\\
We are also in the position to investigate the minimum energy configurations (MEC) of Beltrami carbon pseudospheres as a function of the number of interacting particles, with the explicit inclusion of the electronic degrees of freedom (generalized Thomson problem \cite{Thomson}). Thus, among a number of FIRE optimized pseudospheres with the same curvature and number of atoms we performed first-principle simulations based on the density-functional tight-binding (DFTB) method within the Born-Oppenheimer approximation \cite{frauenheim2002atomistic,garberoglio2012modeling} (details on parameters, convergence issues and DFT against DFTB accuracy checks are provided in the SM). The first notable result is that, releasing the constraint that carbon atoms lay on the Beltrami surface, the pseudosphere containing 1146 carbon atoms is stable at both DFT and DFTB levels of theory.
With respect to the constrained optimization, we find the appearance of bumps in correspondence of the uncharged penta-heptagonal scars, locally spoiling the constant curvature (see Fig. S11 of the SM).\\
Restoring the constraints, we finally performed the total energy ground state calculation of 9 FIRE optimized Beltrami pseudospheres containing 1146 carbon atoms, see SM. In the bottom panel of Fig. \ref{1146} we report the energy landscape. Minima are indeed critically affected by the presence of the external ring due to the different number and position of defects. Thus, we can find MEC only in statistical sense, providing a number of pseudospheres with a comparable energy minimum, excluding those energetically far. In this case, the energy difference per carbon atom between the pseudospheres labelled as 4,5 and 7 is below 0.05 eV, which is of the order of magnitude of the energy difference between different carbon atom bonds. Thus, these three configurations represent the most likely MEC candidates for our statistics, and in the top panel we report these 3 Beltrami structures.
We repeated the same calculations for larger samples, notably having 2146, 5506 carbon atoms to shed some light on the behaviour of the energy minimum with increasing $N$. We report the relevant energy landscape in Fig. \ref{2146en}, while the optimized geometries can be found in the SM (see Figs. S9 and S10 of the SM).

In conclusion, we have realized, for the first time in a simulation lab, a Lobachevsky molecular structure, facing and solving the various issues of embedding  Lobachevsky geometry in $\mathbb{R}^3$. We have found: i) a specific NEC group for Beltrami; ii) how to face the Hilbert horizon; iii) a novel mechanically stable carbon geometry. This leads to conjecture that the infinitely many surfaces of constant negative $K$ could correspond to the infinitely many NEC groups, suitably adapted to $\mathbb{R}^3$ (e.g., a natural candidate for a non-cyclic, infinite order generalization of $F_Y$ is the Dini surface \cite{solitonANDdini}). This work paves the way to the realization of systems corresponding to QFT in (quantum) gravitational backgrounds \cite{iorioreview}, hence, e.g., to test black hole quantum physics \cite{iorio}.

\begin{acknowledgments}
A.~I. acknowledges the Czech Science Foundation (GA\v{C}R), Contract No. 14-07983S, for support, and thanks Fondazione Bruno Kessler (Trento)
for the hospitality. S.T and N.M.P. acknowledge funding from the EU under the FP7th grant agreement 604391 (Graphene Flagship) and from the Provincia Autonoma di Trento (``Graphene nanocomposites'', no. S116/2012-242637 and reg. delib. no. 2266). N.M.P. is also supported by the European Research Council (ERC StG Ideas 2011 BIHSNAM no. 279985 on 'Bio-inspired hierarchical supernanomaterials', ERC PoC 2013-1 REPLICA2 no. 619448 on 'Largearea replication of biological anti-adhesive nanosurfaces', ERC PoC 2013-2 KNOTOUGH no. 632277 on 'Super-tough knotted fibres').

\end{acknowledgments}

\newpage

\centerline{\Huge Supplemental Material}

\vskip 1cm

\section{Lobachevsky geometry in $\mathbb{R}^3$}

The upper-half plane, $\{ (\tilde{x}, \tilde{y}) | \tilde{y} > 0 \}$, equipped with the metric
\begin{equation}
d l^2 = \frac{r^2}{{\tilde y}^2}(d{\tilde x}^2+d{\tilde y}^2) \label{lobsurface}\;,
\end{equation}
represents Lobachevsky geometry, both locally and globally. The geodesics for (\ref{lobsurface}) are semi-circles, starting and ending on the ``absolute'', that is the $\tilde{x}$-axis. These include the limiting case of infinite radii semicircles, that are straight half-lines parallel to $\tilde{y}$. To realize this geometry in a real laboratory we need to embed (\ref{lobsurface}) in $\mathbb{R}^3$, that is, we need to find a surface with constant $K=-r^{-2}$. This means to specify ${\tilde x}$ and ${\tilde y}$ in terms of coordinates measurable using the Euclidean distance (embedding), for instance the parallel and meridian coordinate, $(u,v)$, respectively (see later).

Each such surface is only {\it locally} isometric to the Lobachevsky plane, and only represents a ``stripe'' (a sector of the area between two parallels, called ``horocyclic sector'' [3]). Since there are infinitely many ways to cut-off a stripe from the Lobachevsky plane, it is easy to convince oneself that there are infinitely many such surfaces. Furthermore, each carries some sort of singularity: cusps, self-intersections, boundaries, etc. These are unavoidable, as proved in the Hilbert theorem stated in the main text. Intuitively, we might say that these singularities/infinities stem from forcing an intrinsically non-compact structure (the isometries of (\ref{lobsurface}) form the group $SL(2,\mathbb{R})$, locally isomorphic to $SO(2,1)$, the Lorentz group in 2+1 dimensions) into a space, $\mathbb{R}^3$, whose isometries are given by the compact group $SO(3)$. Indeed, the Hilbert theorem does not apply to embeddings into $\mathbb{R}^{(2,1)}$.

We show here how this comes about in the case of the Beltrami pseudosphere of interest, that is one of the surfaces of constant negative $K$ that is also a \textit{surface of revolution}, like the sphere. This will illustrate, among other things, the nature of the singular boundary that we call ``Hilbert horizon'', in honor of Hilbert.

\begin{figure}
\centering
%\leavevmode \epsfxsize=10cm \epsfysize=10cm
\includegraphics[height=.4\textheight]{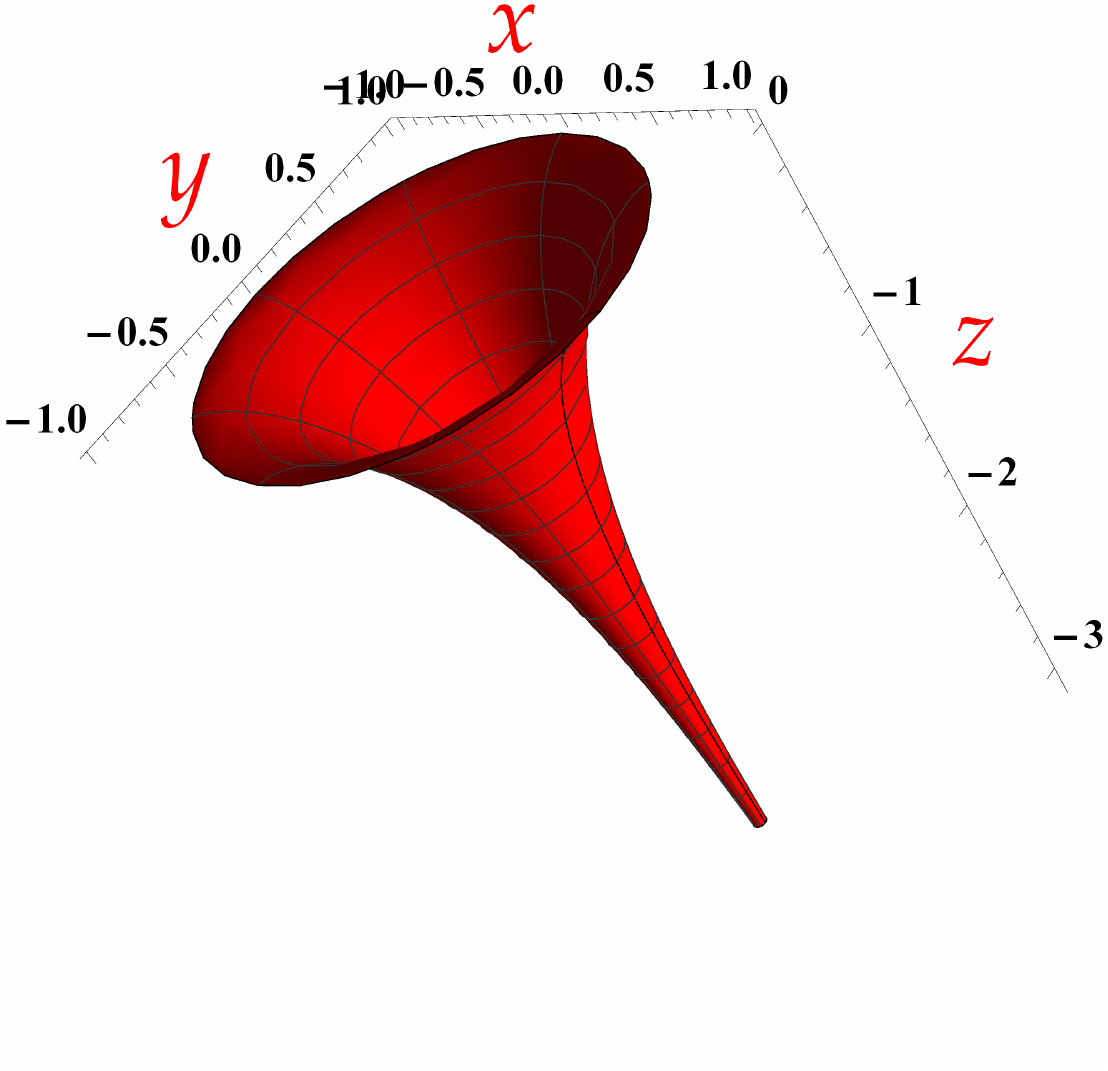}
%\epsffile
\caption{The Beltrami pseudosphere, plotted for $r=1=c$ and $u \in [-3.4, 0]$, $v\in [0, 2\pi]$.} \label{Beltrami}
\end{figure}

A surface of revolution is a surface swapped by a profile curve, say in the plane $(x,z)$, rotated of a full angle around the $z$-axis. All such surfaces can be parameterized as [18]
\begin{equation} \label{canonicalpar}
x(u,v) = R(u) \cos v \;, \; y(u,v) = R(u) \sin v \; , \; z(u) = \pm \int^u \sqrt{1 - {R'}^2(\bar{u})} d \bar{u} \;,
\end{equation}
where prime denotes derivative with respect to the argument, $v\in [0, 2 \pi]$ is the parallel coordinate (angle), and $u$ is the meridian coordinate. The range of $u$ is fixed by the knowledge of $R(u)$, i.e. of the type of surface, through the request that
\begin{equation}\label{constraint}
z (u) \in \mathbb{R} \;.
\end{equation}

When $K$ is constant we have
\begin{equation}\label{solgaussian1}
R(u) = c \cos (u/r + b) \quad {\rm for} \quad K = \frac{1}{r^2} > 0 \;,
\end{equation}
and
\begin{equation}\label{solgaussian2}
R(u) = c_1 \sinh (u/r) + c_2 \cosh (u/r )\quad {\rm for} \quad K = - \frac{1}{r^2} < 0 \;,
\end{equation}
where $r, c, b, c_1, c_2$ are real constants, that determine the type of surface, and/or set the zero and scale of the coordinates.

For $K = 1/r^2$, one first chooses the zero of $u$ in such a way that $b=0$, then distinguishes three cases:
$c = r$ (sphere), $c < r$ (spindle), $c > r$. With these,
\begin{equation} \label{zkpositive}
z(u) = \int^u \sqrt{1 - (c^2 / r^2) \sin^2(\bar{u}/r)} d\bar{u} \;.
\end{equation}
This elliptic integral has singularities when $c \neq r$, but these cases are applicable to the sphere, $c=r$, through a simple redefinition of the meridian coordinate $v \to (c/r) v$. Therefore, these singularities are inessential and easily avoidable. Once that is done, the integral gives $z(u) = r \sin (u/r)$, with $u/r \in [-\pi/2 , + \pi/2]$.

For $K = - 1/r^2$, all the surfaces described by (\ref{solgaussian2}) can be applied to one of the following three cases: either $c_1 = c_2 \equiv c$, giving
\begin{equation} \label{Rbeltrami}
R(u) = c \; e^{u/r} \;,
\end{equation}
or $c_1 \equiv c$, $c_2 = 0$, giving
\begin{equation} \label{Relliptic}
R(u) = c \; \sinh (u/r) \;,
\end{equation}
or $c_1 = 0$, $c_2 \equiv c$, giving
\begin{equation} \label{Rhyperbolic}
R(u) = c \; \cosh(u/r) \;.
\end{equation}
They are called the \textit{Beltrami}, the \textit{elliptic}, and the \textit{hyperbolic} pseudospheres, respectively. Notice that for the Beltrami and hyperbolic surfaces, $c$ is only bound to be a real positive number. For the elliptic surface, instead, $0 < c \equiv r \sin \beta < r$, where $\beta$ is the angle between the axis of revolution, and the tangents to the meridians at $R=0$ (see later).

The corresponding expressions for $z(u)$ are obtained by substituting $R(u)$ in the integral in (\ref{canonicalpar}). Having done that, the condition (\ref{constraint}) gives the range of $u$ (hence of $R$) in the three cases
\begin{eqnarray}
R(u) & \in & [0, r] \quad (u \in [- \infty, r \ln(r/c)]) \;,\label{raggiobel} \\
R(u) & \in & [0, r \cos \beta] \quad (u \in [0 , {\rm arcsinh} \cot \beta]) \;, \label{raggioell}\\
R(u) & \in & [c, \sqrt{r^2 + c^2}] \quad (u \in [- {\rm arccosh} (\sqrt{1 + r^2 /c^2}) , + {\rm arccosh} (\sqrt{1 + r^2/c^2})]) \;,  \label{raggiohyp}
\end{eqnarray}
respectively.

The key difference with $K>0$, and manifestation of the Hilbert theorem for the surfaces of revolution, is that, no matter the redefinition of coordinates, the singularities of the elliptic integral for $z(u)$ can never be avoided.

Summarizing, the Beltrami pseudosphere has coordinates (see (\ref{canonicalpar}))
\begin{equation}
x(u,v) = c \, e^{u/r} \cos v \,, \; y(u,v) = c \, e^{u/r} \sin v \,, \;
z(u) = r (\sqrt{1 - (c^2/r^2) e^{2 u/r}} - {\rm arctanh}\sqrt{1 - (c^2/r^2) e^{2 u/r}}) \;,
\end{equation}
and is identified by $R(u) =  c \, e^{u/r} \in [0,r]$ as $u \in [-\infty, r \ln(r/c)]$. The surface is not defined for $R>r$  as $z(u)$ becomes imaginary. The singular boundary (Hilbert horizon) is the circle $R=r$ (at $u=r \ln(r/c)$). In the main text, for simplicity, we choose $c=r$, that gives $u = 0$ as the parallel coordinate of the Hilbert horizon.

Comparing (\ref{Rbeltrami}) with (\ref{Relliptic}), and (\ref{raggiobel}) with (\ref{raggioell}), one sees that, for $c/r \to 0$ (that is $\beta \to 0$), the Beltrami pseudosphere is a limiting case of the elliptic surface [12].

\section{``Hyperbolizations'' of the sphere and the $F_Y$ group}

Here we show how the deformation of the sphere that produces a Beltrami pseudosphere guides us in the construction of the NEC group $F_Y$ of the main text.

\begin{figure}[h]\label{twohyperbolizations}
\centering
\includegraphics[width=.6\linewidth]{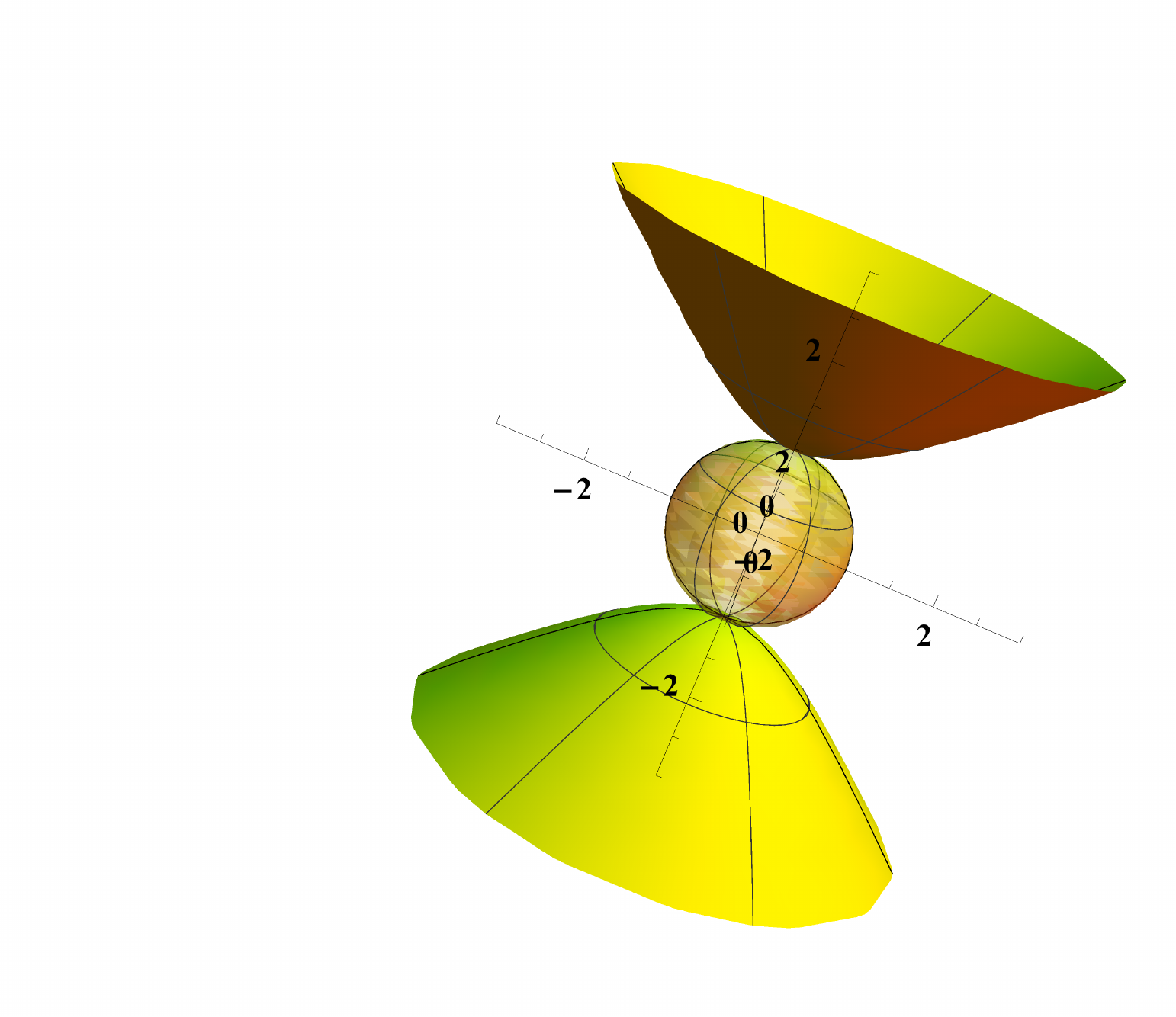}
\includegraphics[width=.6\linewidth]{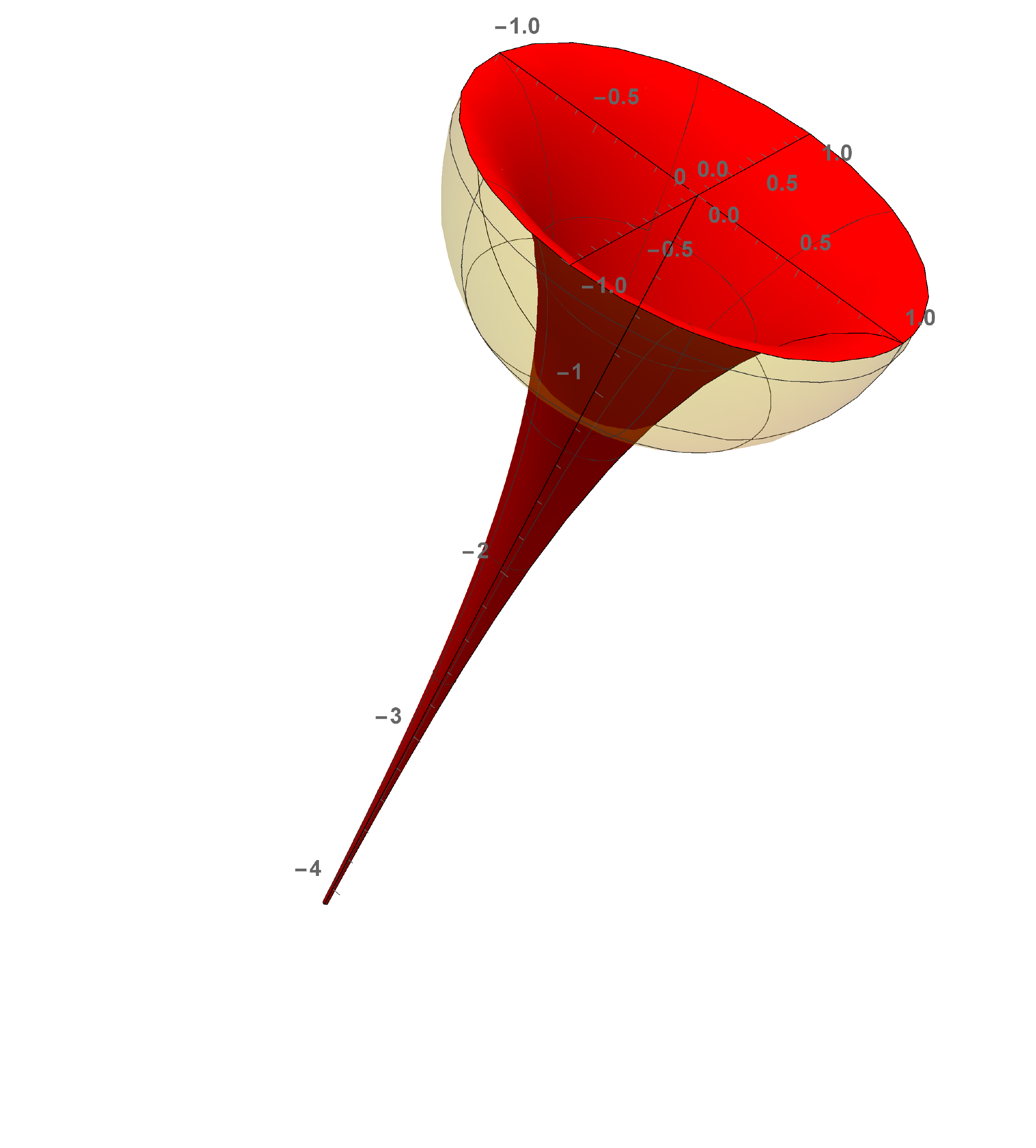}
\caption{A unit sphere in $\mathbb{R}^3$, next to two ``hyperbolizations''. Only the second (the Beltrami pseudosphere) is a realization of Lobachevsky geometry.}
\end{figure}

\begin{figure}[h]\label{anglesconstruction}
\centering
\includegraphics[width=.4\linewidth]{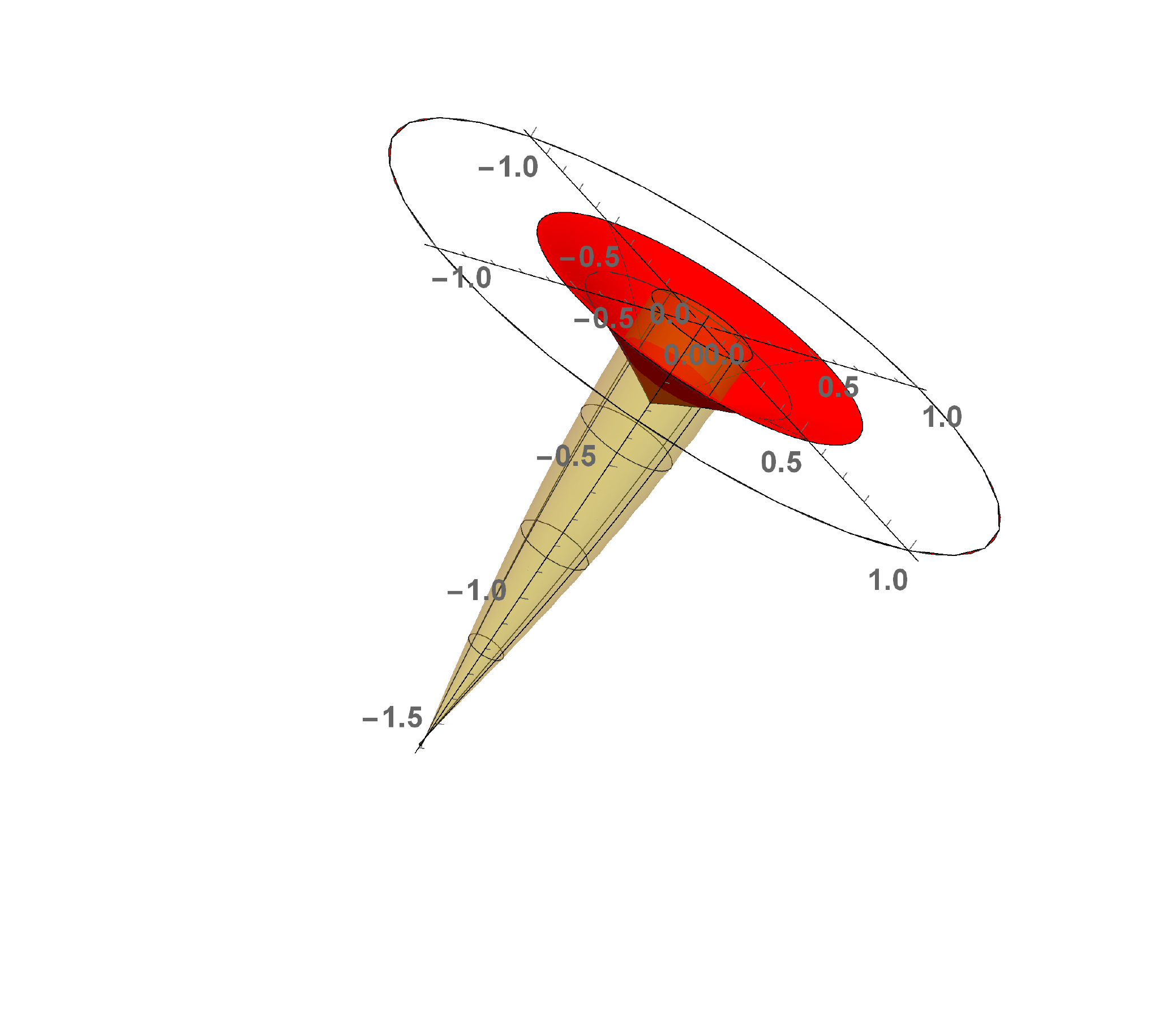}
\includegraphics[width=.4\linewidth]{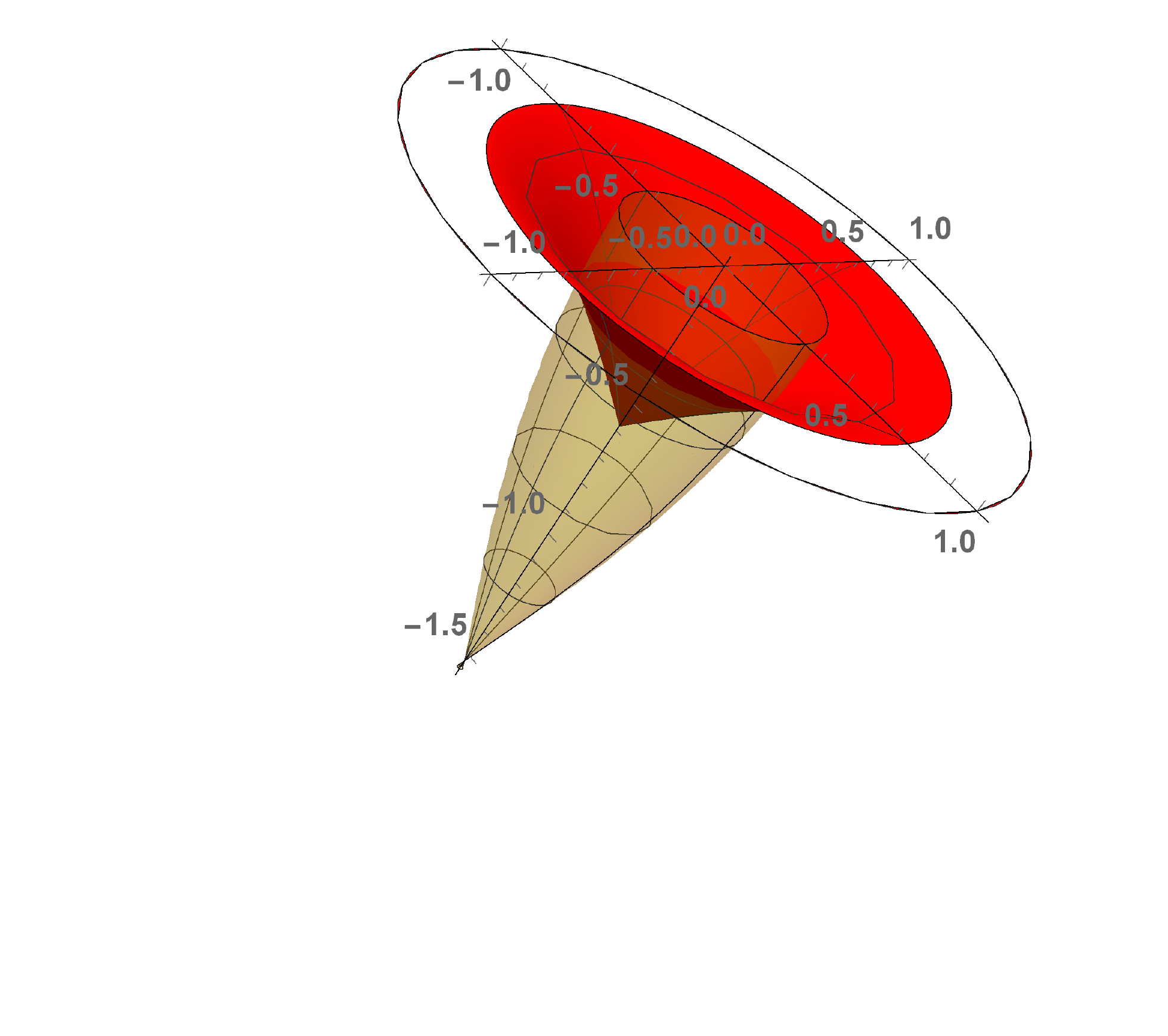}
\includegraphics[width=.4\linewidth]{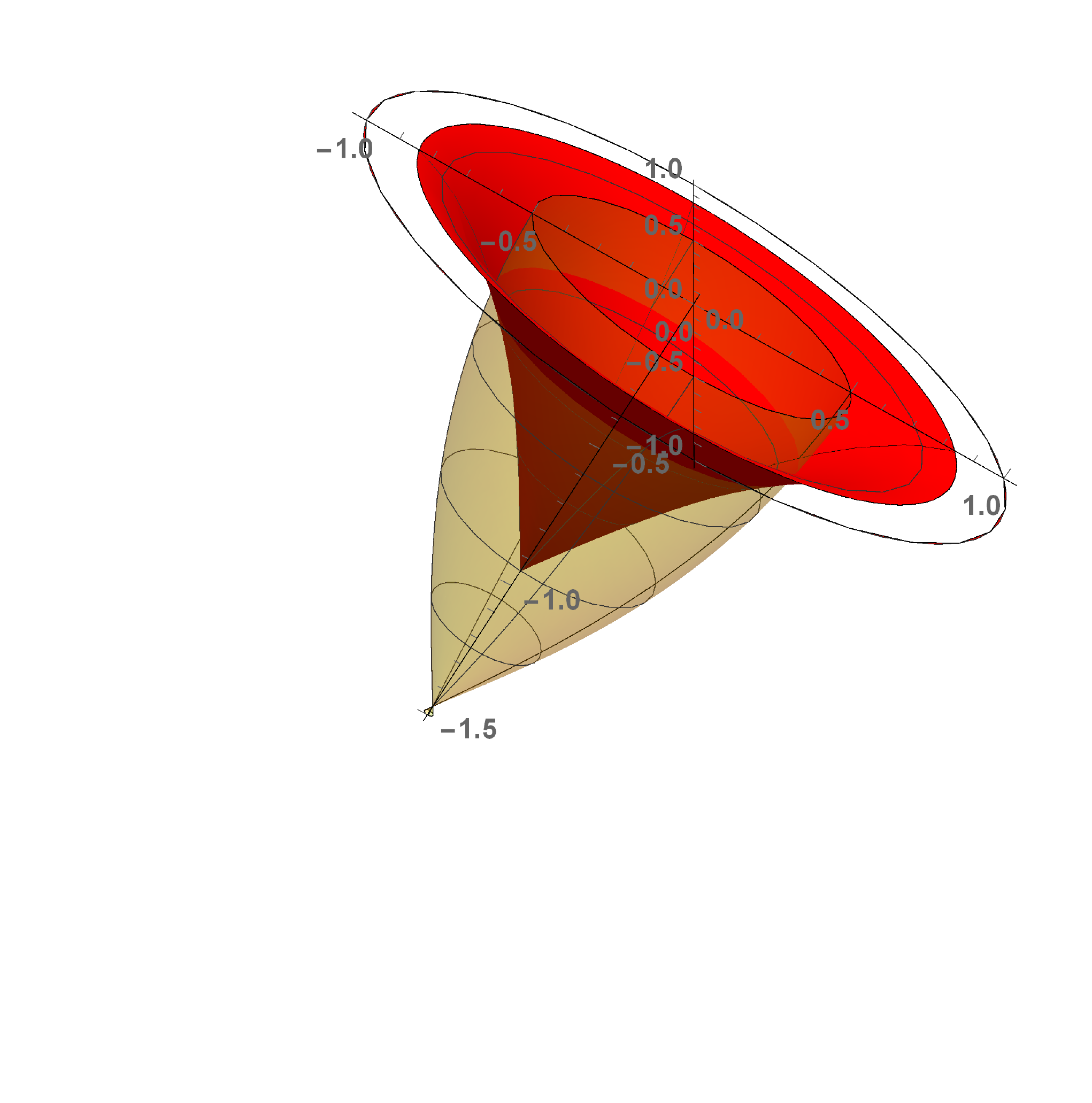}
\includegraphics[width=.4\linewidth]{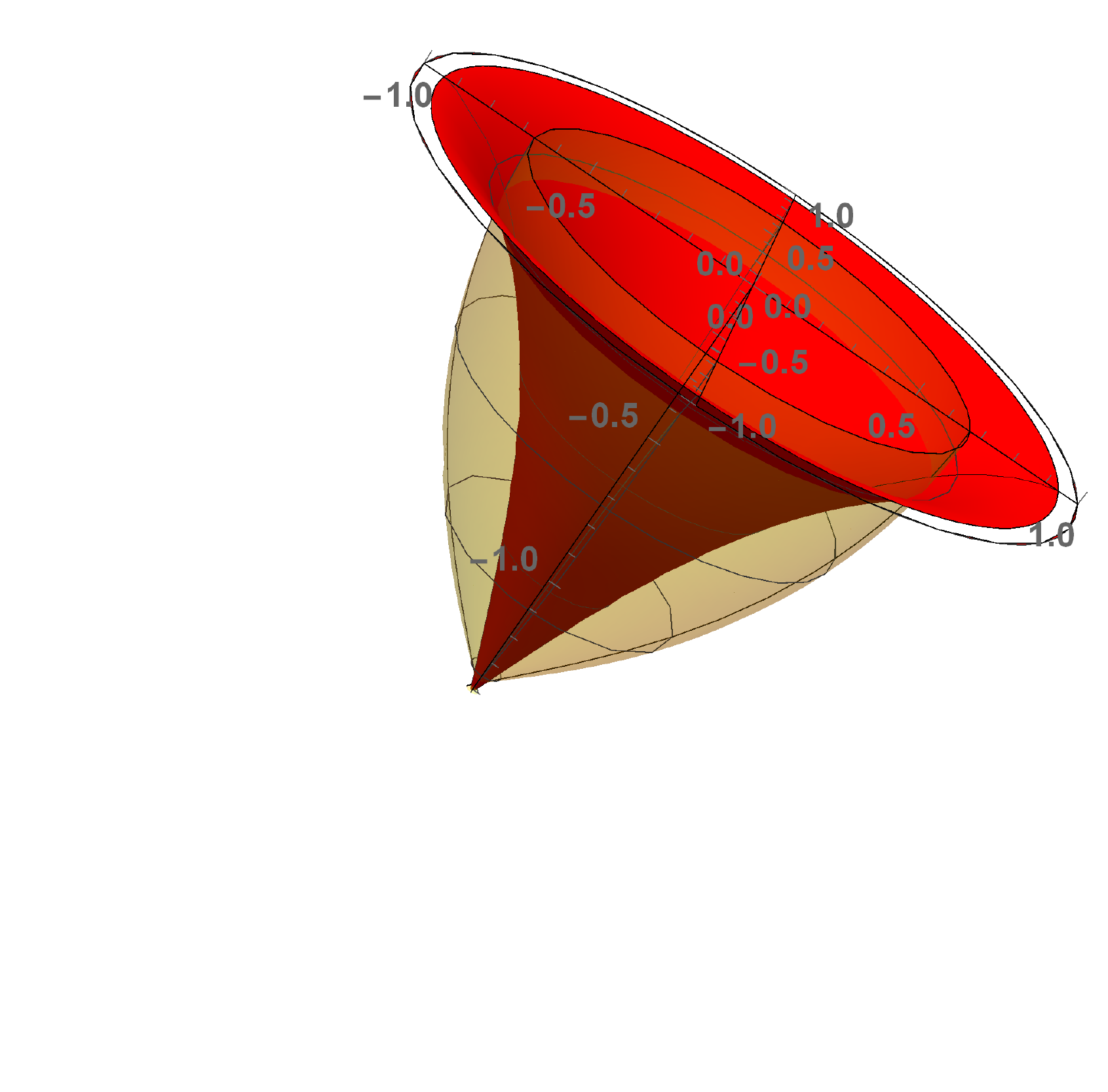}
\includegraphics[width=.4\linewidth]{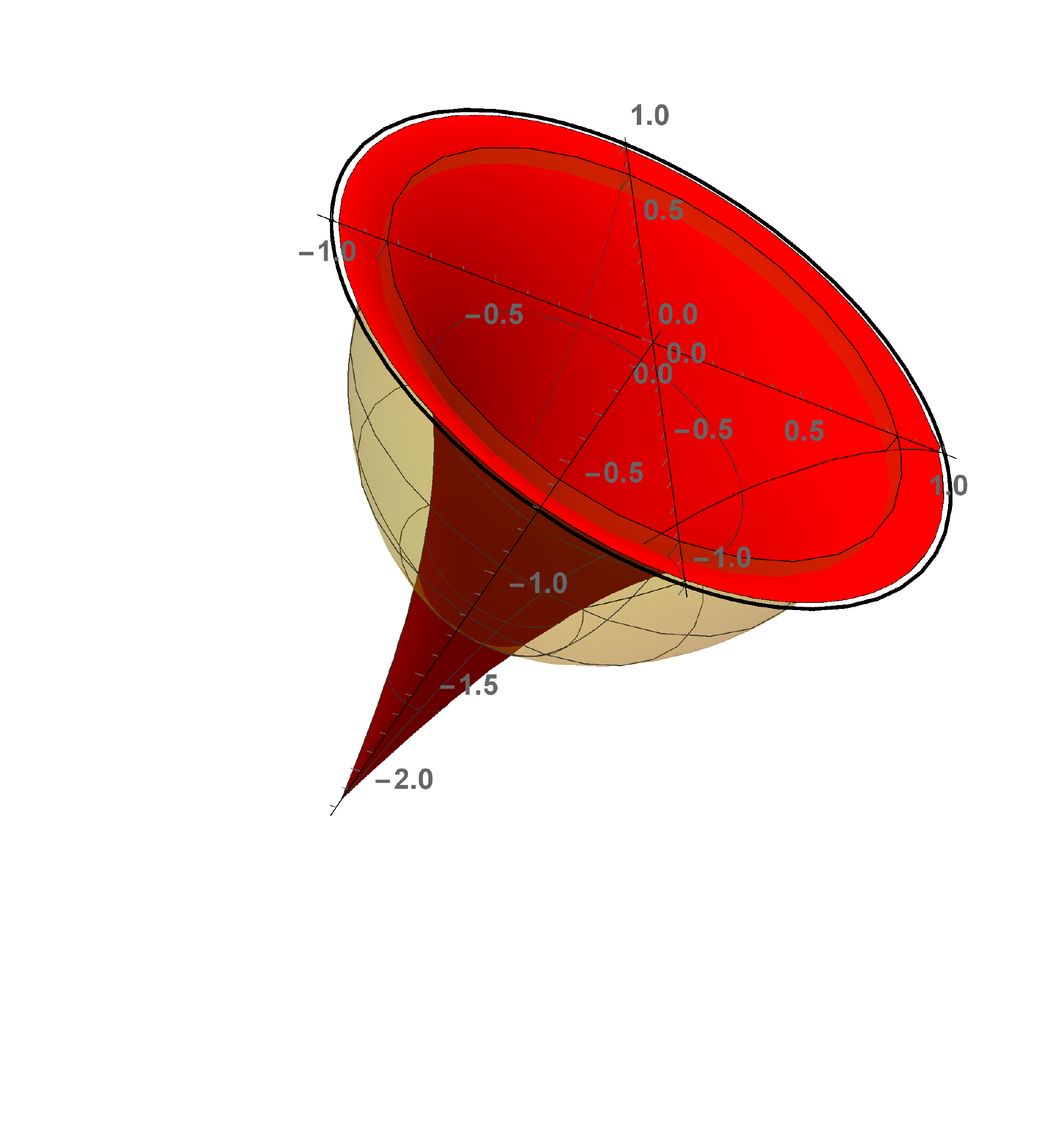}
\caption{Elliptic pseudospheres of total curvatures $K_{tot} = - \pi/3,- 2 \pi/3,- 3 \pi/3,- 4 \pi/3, - 5 \pi/3$ (left to right, top to bottom, respectively), next to the corresponding surfaces of opposite curvatures (half-spindles), showing the process of deformation of one surface into the other (a positive curvature defect at the tip turning into a negative curvature defect at the tip). The limiting cases of $K_{tot} = \mp 6 \pi/3 = \mp 2 \pi$, that are the Beltrami and the half sphere, respectively, are shown in the previous Figure.}
\end{figure}

Consider the equation for the sphere in $\mathbb{R}^3$
\begin{equation}\label{dSsphere}
 x^2 + y^2 + z^2 \equiv x^i \delta_{i j} x^j = + r^2 \;,
\end{equation}
where $x^i \equiv (x,y,z)$. On the left side we have the length squared of the position vector in $\mathbb{R}^3$, and the metric of the latter is $\delta_{i j}= {\rm diag} (+1, +1, +1)$. The symmetry group of the equation is $SO(3)$, as it is seen from the solution of (\ref{dSsphere})
\begin{equation}
x = r \cos v \sin (u/r) \,, \; y = r \sin v \sin (u/r) \,, \; z = r \cos (u/r) \;,
\end{equation}
that shows two compact rotations, one for the parallels ($v$), one for the meridians ($u$). $SO(3)$ is {\it also} the group of isometries of the embedding space $\mathbb{R}^3$. Therefore, the surface identified by such coordinates (the sphere) realizes the geometry of constant positive $K$, and discrete subgroups of $SO(3)$ are related to its tiling, as well as to the crystallographic groups, as recalled in the main text (see also [4]).

Consider now
\begin{equation}\label{AdSsphere}
x^2 + y^2 - z^2 \equiv x^i \eta_{i j} x^j = - r^2 \;,
\end{equation}
where $\eta_{i j} = {\rm diag} (+1, +1, -1)$. The symmetries of this equation form the non-compact group $SO(2,1) \sim SL(2, \mathbb{R})$, where besides standard compact rotations, there are hyperbolic rotations. As before, this can be seen from the solution of (\ref{AdSsphere})
\begin{equation} \label{AdScoordinates}
x = r \cos v \sinh (u/r) \,, \; y = r \sin v \sinh (u/r) \,, \; z = r \cosh (u/r) \;,
\end{equation}
where the rotation for the meridians ($u$) becomes hyperbolic. $SL(2,\mathbb{R})$ is the isometry group of Lobachevsky geometry, whose discrete subgroups are, by definition, the NEC groups.

If the embedding space for (\ref{AdSsphere}) is $\mathbb{R}^3$, the symmetries of the equation and of the embedding space \textit{differ}, being $SO(3)$ for the latter, and $SL(2,\mathbb{R})$ for the former. In fact, in $\mathbb{R}^3$ the surface identified by the coordinates (\ref{AdScoordinates}) is a double-sheet hyperboloid, showed in Fig.~2, that is a surface of non-constant positive Gaussian curvature, $K = [r \cosh(2u)]^{-2} > 0$, that nothing has to do with Lobachevsky geometry.

If the embedding space for (\ref{AdSsphere}) is $\mathbb{R}^{(2,1)}$, the symmetries of the equation and of the embedding space \textit{coincide}, and indeed we have a realization of Lobachevsky geometry. This can be seen by writing the line element
\begin{equation} \label{ellpseudospherer21}
dl^2 \equiv dx^2 + dy^2 - dz^2 = du^2 + (r \sin \beta)^2 \sinh^2 (u/r) dv^2 \;,
\end{equation}
where $x$,$y$, and $z$ are given by (\ref{AdScoordinates}), but with the parameter $c=r \sin \beta$ reintroduced (see previous Section). Indeed, this is the line element of the elliptic surface of the previous Section. If the embedding space is only artificial, i.e. the only physical space is the two-dimensional surface, then the line element in terms of $u$ and $v$ in (\ref{ellpseudospherer21}) is all that is necessary. If the embedding space is physical and is $\mathbb{R}^3$, like for us, then the coordinates that one needs to use cannot be (\ref{AdScoordinates}), but (see (\ref{canonicalpar}) and (\ref{Relliptic}))
\begin{equation} \label{ellipticxy}
x = (r \sin \beta) \cos v \sinh (u/r) \;, \; y = (r \sin \beta) \sin v \sinh (u/r) \;,
\end{equation}
that are very similar to $x$ and $y$ in (\ref{AdScoordinates}), but the $z$ coordinate is a complicated expression given in terms of elliptic integrals of the E and F type
\begin{eqnarray} \label{ellipticz}
z & = & 2 r \cot^2 \beta {\rm csch}(\frac{2 u}{r}) {\Big(} {\rm EllipticE}
{\Big [} \arcsin(\frac{1}{2} \sqrt{3 + \cos(2 \beta) - 2 \cosh(\frac{2 u}{r}) \sin^2 \beta}), \sec^2 \beta {\Big ]} \nonumber \\
& - &  {\rm EllipticF} {\Big [} \arcsin(\frac{1}{2} \sqrt{3 + \cos(2 \beta) - 2 \cosh(\frac{2 u}{r}) \sin^2 \beta}),
    \sec^2 \beta {\Big ]} {\Big)} \sqrt{\cosh^2(\frac{u}{r}) \sin^2 \beta} \sqrt{ \sinh^2(\frac{u}{r}) \tan^2 \beta} \;.
\end{eqnarray}
Indeed, when we write $dl^2 \equiv dx^2 + dy^2 + dz^2$ and use the coordinates (\ref{ellipticxy}) and (\ref{ellipticz}), we obtain
\begin{equation} \label{ellpseudospherer3}
dl^2 \equiv dx^2 + dy^2 + dz^2 = du^2 + (r \sin \beta)^2 \sinh^2 (u/r) dv^2 \;,
\end{equation}
where we recognize, on the far-right side, the same line element obtained in (\ref{ellpseudospherer21}), hence, we have realized  Lobachevsky geometry in $\mathbb{R}^3$. This is not yet our Beltrami pseudosphere, that is reached in the limit $c/r \sim \beta \to 0$. This correspondence is of much help to find $F_Y$, as we now show.

First we see that the area of the elliptic pseudosphere is
\begin{equation}
A^E = 2 \pi \int_{u_{min}}^{u_{max}} R(u) = 2 \pi c  \int_{u_{min}}^{u_{max}} \sinh(u/r) = 2 \pi r^2 (1 - \sin \beta) \;,
\end{equation}
reaching the area of the Beltrami, $A^B = 2 \pi r^2$, for $\beta \to 0$. From the Gauss-Bonnet formula of the main text, we know that six heptagonal defects are necessary for the Beltrami, $K_{tot} = - 2 \pi = 6 \times (- \pi/3)$. Therefore, being $A^E \leq A^B$, we can thing of progressively accommodating one unit of negative curvature $(- \pi/3)$ at the time, from 1 to 6. This way we shall form elliptic pseudospheres whose areas are such that $K_{tot} = - \pi/3$, then $K_{tot} = - 2 \pi/3$, and so on, till $K_{tot} = - 5 \pi/3$, and then obtain the case of $K_{tot} = - 6 \pi/3 = - 2 \pi$ as a limiting case corresponding to Beltrami. In formulae, for one unit of negative curvature
\begin{equation}
A_1^E \equiv \frac{1}{6} A^B \quad {\rm or} \quad 2 \pi r^2 (1 - \sin \beta_1) =  \frac{1}{6} 2 \pi r^2 \;,
\end{equation}
that gives $\beta_1 = \arcsin (5/6)$.

For two units
\begin{equation}
A_2^E \equiv \frac{2}{6} A^B \quad {\rm or} \quad 2 \pi r^2 (1 - \sin \beta_2) = \frac{2}{6} 2 \pi r^2 \;,
\end{equation}
that gives $\beta_2 = \arcsin (4/6)$, and so on. Then, the six angles between the axis of revolution and the tangents to the meridians at $R=0$, are
\begin{equation}
\beta_k = \arcsin[(6-k)/6] \;,
\end{equation}
with $k = 1,..., 6$, where $k=6$ is there only in the limiting case of Beltrami. As will be clear from what follows, these angles are the $\alpha_n$ of the text
\begin{equation}
\beta_k = \arcsin[(6-k)/6] \equiv \alpha_{6 -k} \equiv \alpha_n \;,
\end{equation}
where $n \equiv (6 - k) \in (0, 1, 2, ..., 5)$.

In Fig.~3 we show these five elliptic pseudospheres, along with the corresponding surfaces of constant positive curvature that carry the units of curvature, but of opposite sign. The line elements of the latter surfaces are
\begin{equation} \label{sphere}
ds_n^2 = du^2 + r \sin \alpha_n \sin^2 (u/r) dv^2 \;,
\end{equation}
with the $\alpha_n$ defined above (they are the (half) spindles introduced in the previous Section).

Let us now construct the $F_Y$ of the text, and let us start from the elliptic surface with one defect. Evidently (see Fig.~4) the defect has to be at the tip, and its work is to turn the cone of angle $\alpha_5$, tangent to the surface at the tip ($R=0$), into the negative curvature surface we want. In Fig.~4 we indicate the defects with circles. The role of the tip, though, is special, as in the limiting case of our interest, this point will go to infinity.

Let us move to the case of two units of curvature. We can either use a single ``charge $-2$'' defect, for which $K_{tot} = - 2/3 \pi$ (for a graphene-made membrane, it corresponds to an octagon), or two defects of charge $-1$ (two heptagons for graphene). We shall use the second option. In this case, the charge $-1$ defect at the tip is not enough to turn the tangent cone of angle $\alpha_4$ into the wanted surface. The action of the second defect of charge $-1$ is necessary. That has to turn the tangent cone of angle $\alpha_5$ into a negative curvature portion of the surface. This cone of angle $\alpha_5$ is tangent to the surface at a different point. Therefore the surface is obtained by the combined action of these two defects. This is illustrated in Fig.~4. These two defects correspond to the location of the vertices of the Lobachevsky polyhedron, and enjoy only a $C_1$ symmetry. This is so because one of them is at the tip, hence on the axis of symmetry. Another way of looking at this construction is to think of the process of going from surface 1 (first plot in Fig.~4) to surface 2 (second plot in Fig.~4), as a transformation that keeps the first defect/vertex at the tangent point with fixed angle $\alpha_5$.

As illustrated in detail in Fig.~4, the process continues in this fashion for growing number of defects: for the surface 3, the deformation keeps the defects/vertices of surface 2 at the tangent points corresponding to their angles $\alpha_4$ and $\alpha_5$, and a new defect at the tip, i.e. located at the tangent point with angle $\alpha_3$, appears, and so on. When the defects become 3 or more the rotation symmetry of the surface of revolution will act on them. Since the defect at the tip lies on the axis, the finite rotations are: $C_2$ for the surface 3, $C_3$ for the surface 4, $C_4$ for the surface 5, and in the limit, $C_5$ for the Beltrami (surface 6). Notice that in Fig.~4 we do not apply the rotation symmetry to the defects/vertices. This is done to show how the defects are distributed along the meridian ($u$) coordinates.

Finally, in Fig.~5, the limiting case of the Beltrami shows the value of $R$ where the defects/vertices sit, corresponding to the values of the discrete hyperbolic $u$-motions (``boost'') given in the text.

\begin{figure}[h]\label{conesfromelliptic}
\centering
\includegraphics[width=.45\linewidth]{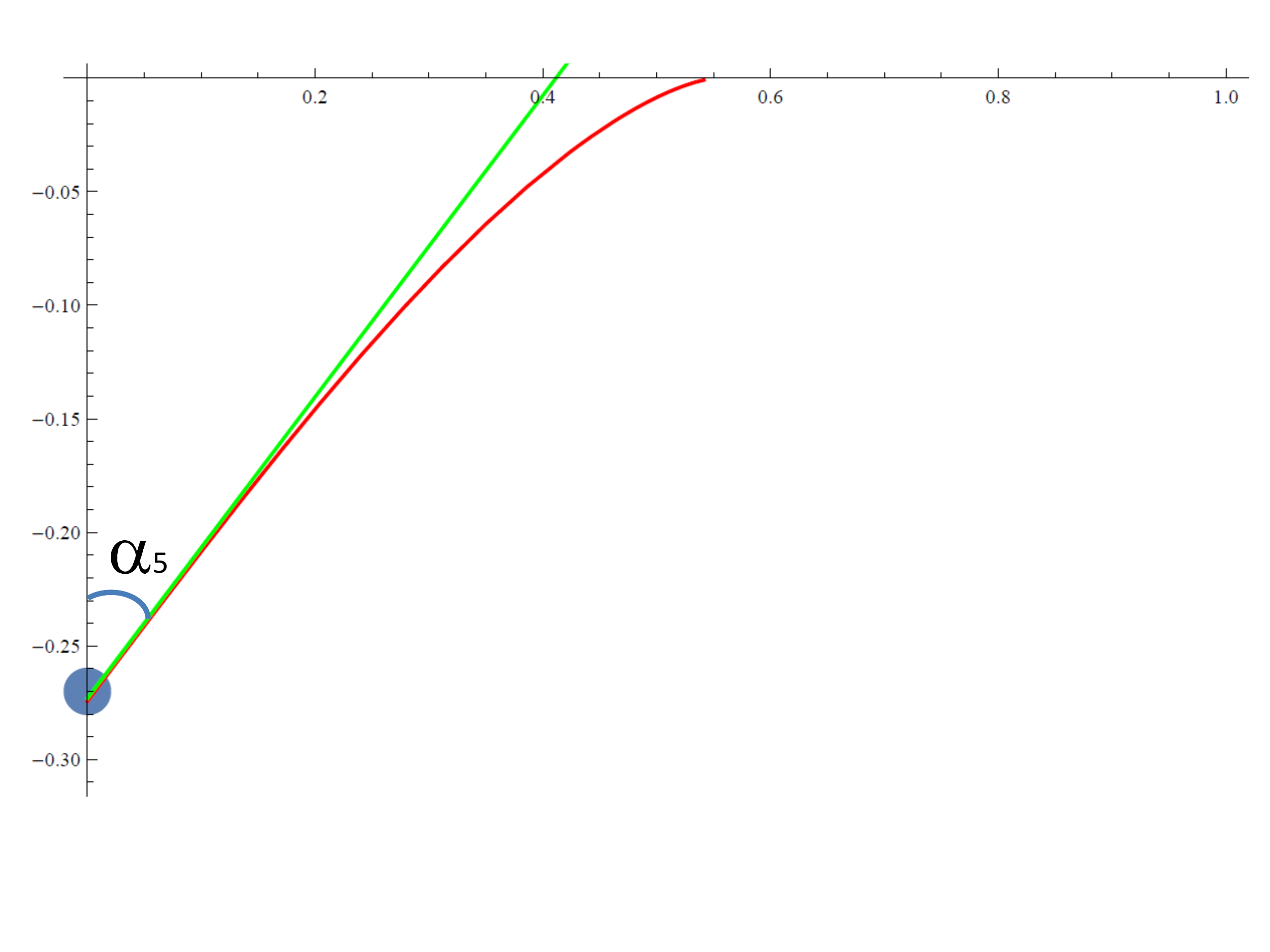}
\includegraphics[width=.45\linewidth]{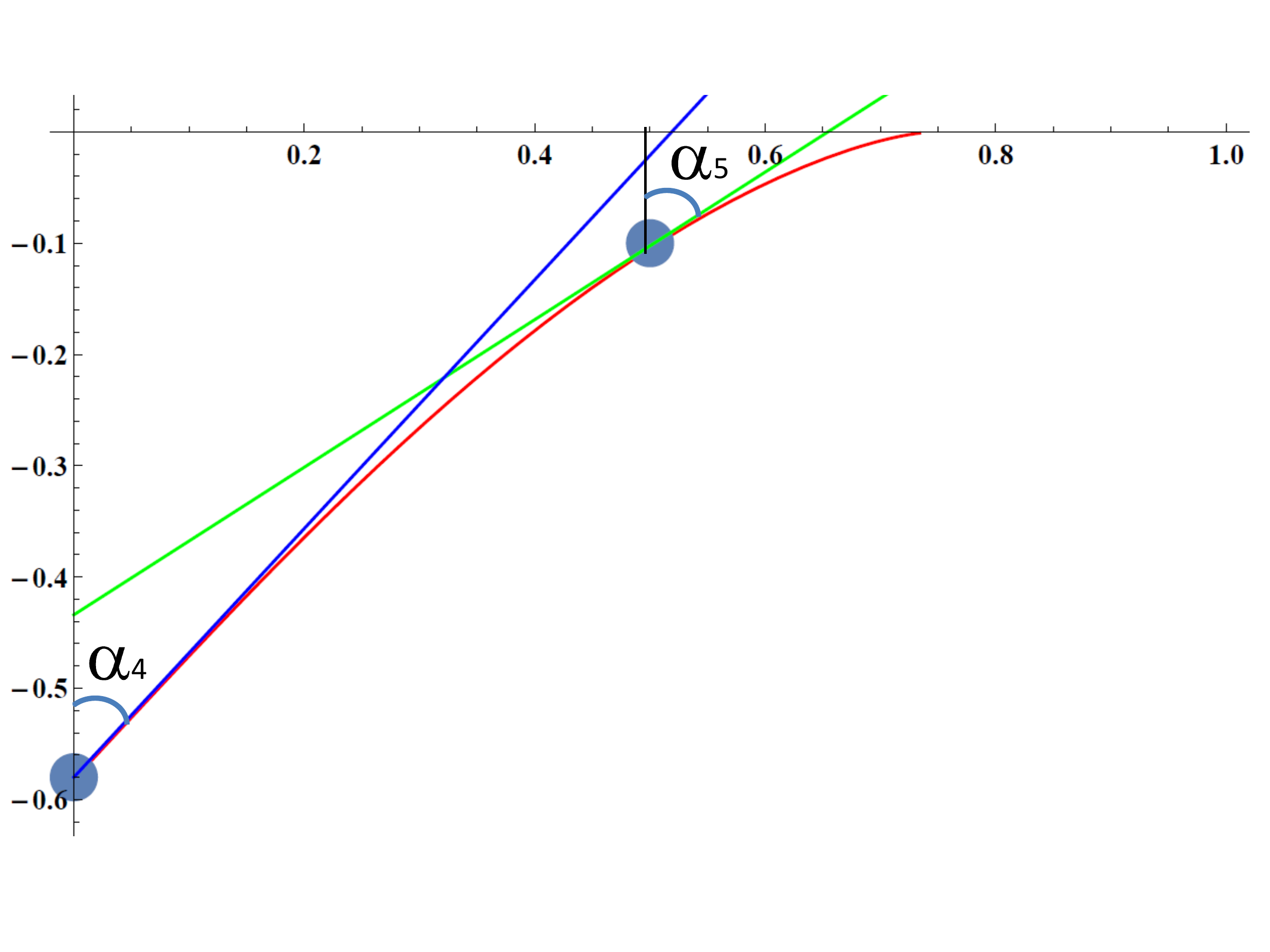}
\includegraphics[width=.3\linewidth]{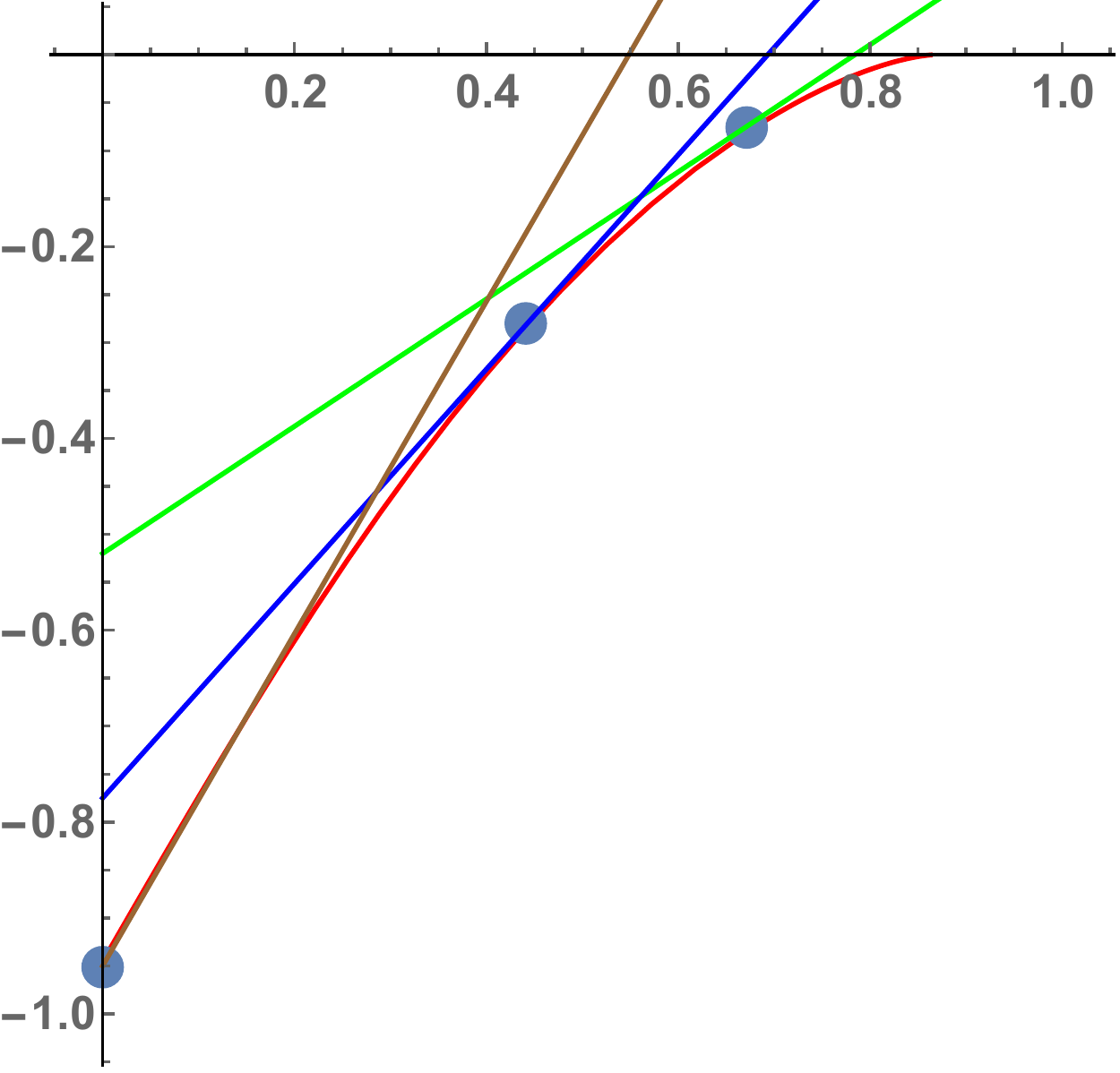}
\includegraphics[width=.3\linewidth]{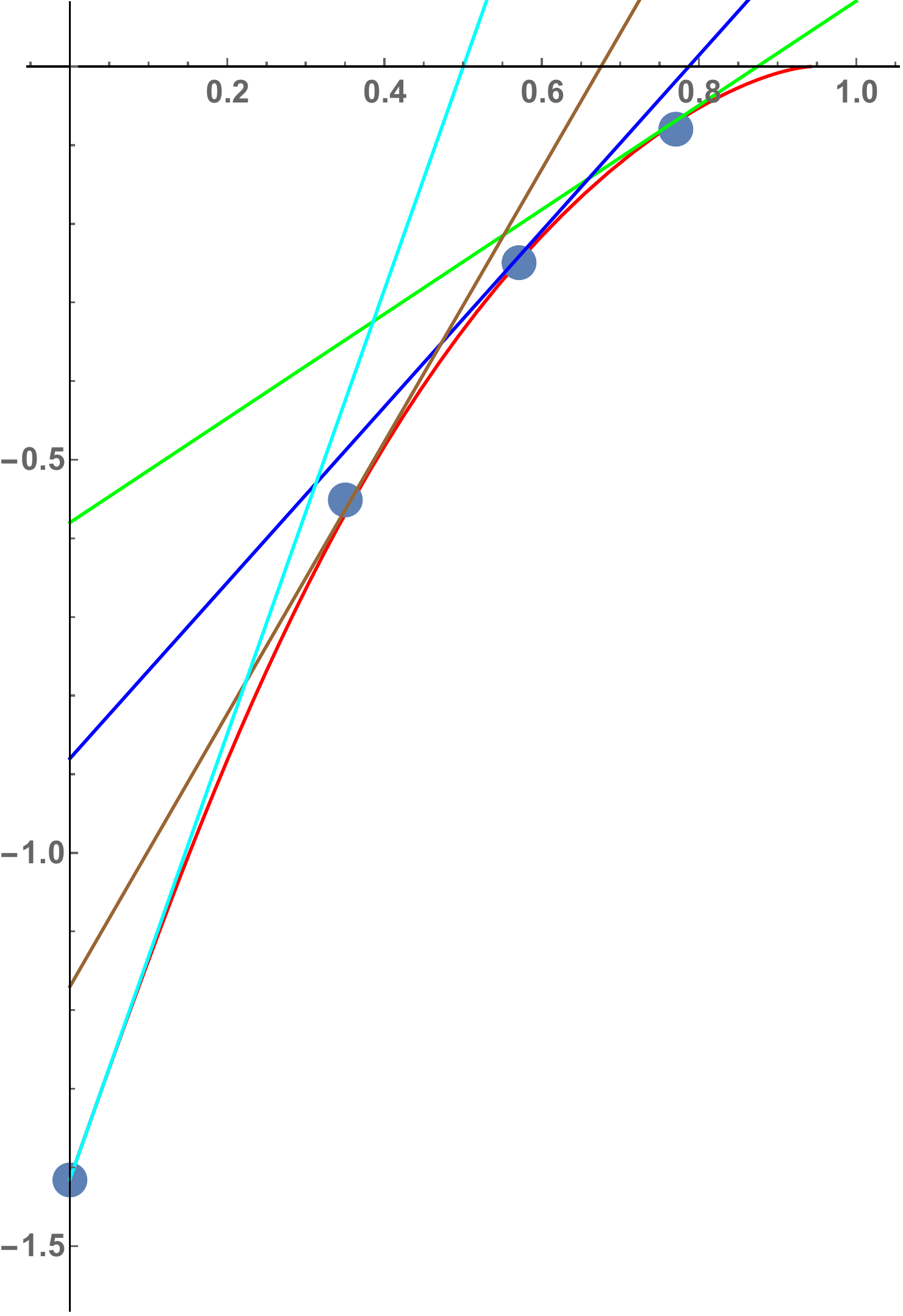}
\includegraphics[width=.3\linewidth]{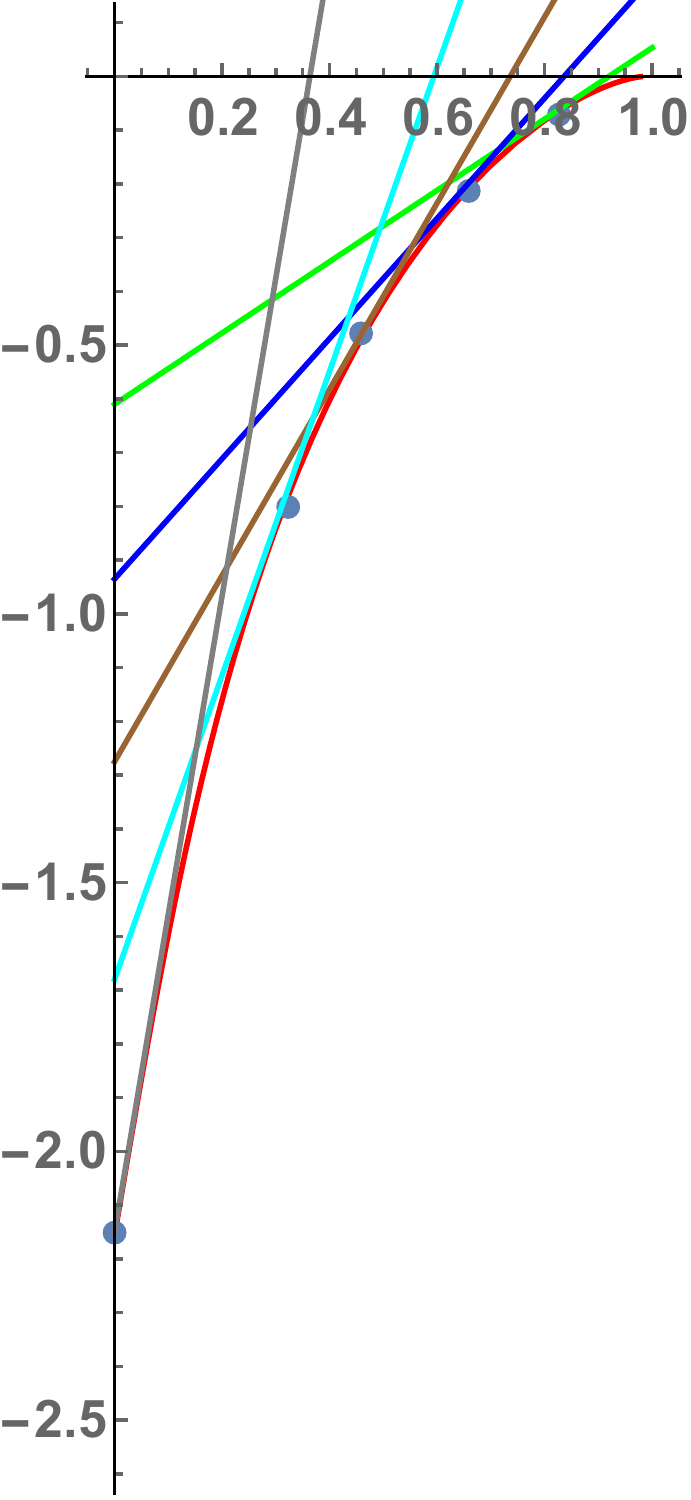}
\caption{Arrangement of charge $-1$ (curvature $K=-\pi/3$) defects for discrete increasing of the negative curvature on the elliptic pseudospheres (see also previous Figure). The first two figures illustrate explicitly how, at each step in the construction, the angles are associated to the defects. The sections of pseudospheres and of cones need be rotated around the $z$-axis (the perpendicular axis here) of $2\pi$. Notice, however, that for the full surfaces this $2\pi$ ($C_1$) rotation must not be applied to the defects/vertices, that only need to be rotated of the angle corresponding to the $C_p$ associated to the given surface, that is: $C_2$ for surface 3 (first left, bottom panel); $C_3$ for surface 4 (middle, bottom panel); $C_4$ for surface 5(last right, bottom panel).}
\end{figure}

\begin{figure}[h]\label{beltramicones}
\centering
\includegraphics[width=.4\linewidth]{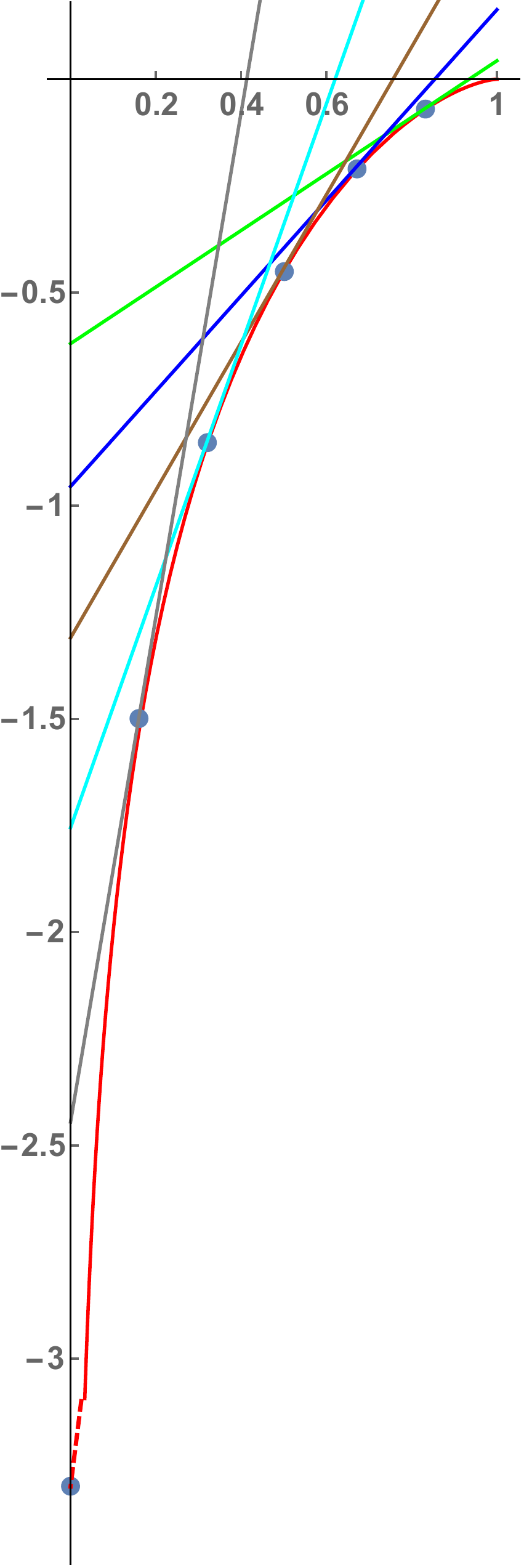}
\caption{Arrangement of the defects/vertices on the Beltrami. The conventions for the plot are the same as those of the previous figure. We also show the value of $R$ corresponding to the location of the defects/vertices, $R/r = 0, 1/6, ..., 5/6$.}
\end{figure}

\section{FIRE with Lennard-Jones potential}

\begin{figure}[h]
\centering
\includegraphics[width=.7\linewidth]{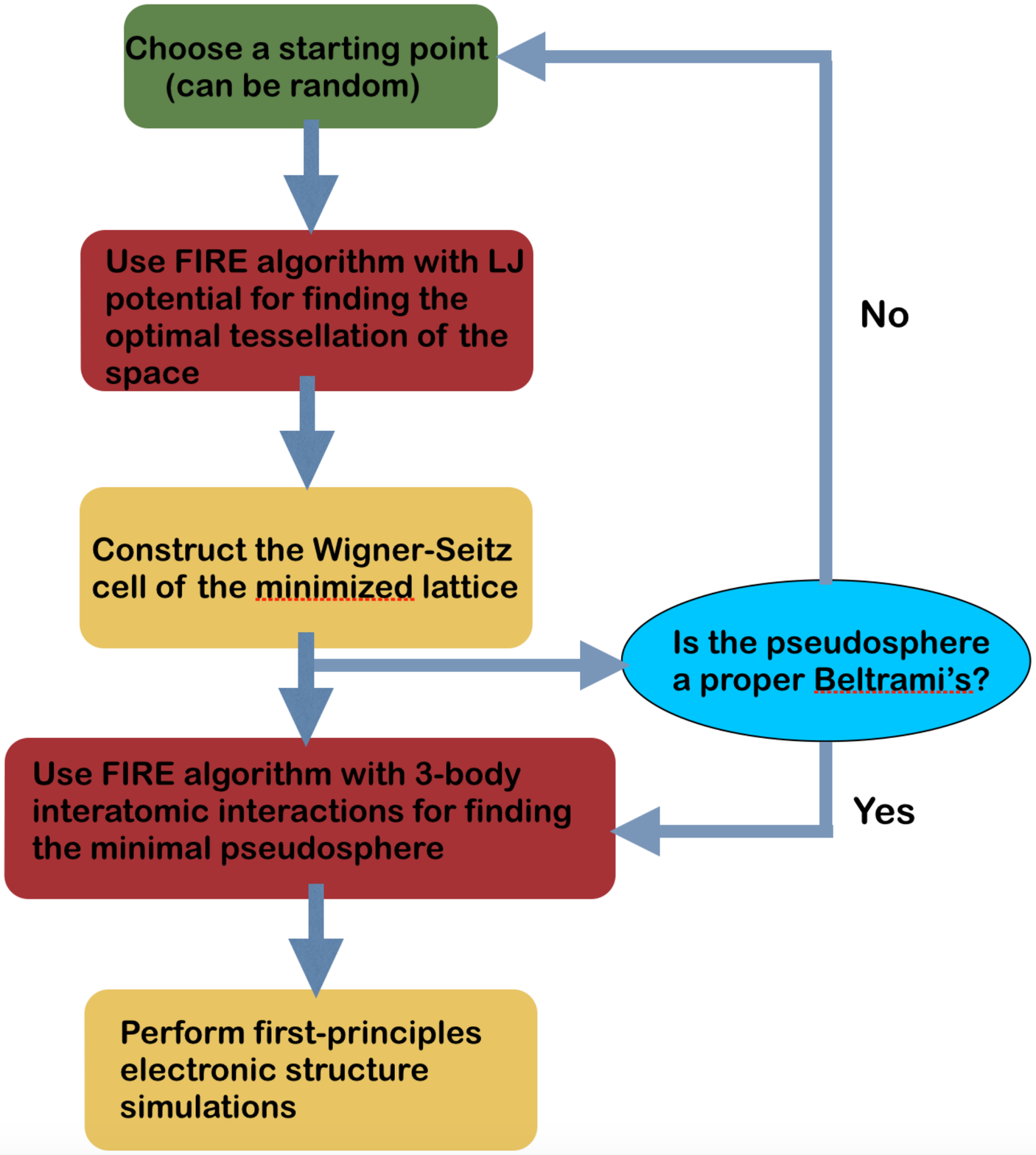}
 \caption{\label{flow} Pseudo-code of our implementation to generate Beltrami psudospheres of minimum energy}
 \end{figure}

Beltrami pseudospheres were generated according to the pseudocode flow  reported in Fig. \ref{flow}. Particles, whose number was chosen so to not exceed that required to cover the entire surface, were initially randomly distributed on the surface and their potential energy was minimized.

Points on the Beltrami's surface interact through a pair-wise Lennard-Jones (LJ) potential as follows:
\begin{equation}\label{LJ}
\phi(r)=4\epsilon \left( \left(  \frac{\sigma^2}{r^2+r_c^2}\right) ^{6} -  \left(\frac{\sigma^2}{r^2+r_c^2}\right)^{3}    \right)
\end{equation}
such that $\phi(2.780 {\text{~ \AA}})=0$, $\phi(r_m)=170$ eV and where $r_c=1.418 {\text{~ \AA}}$ is the potential cut-off radius at short distance.
Potential is set to zero at twice the C-C equilibrium distance of graphene, as after this LJ triangulation (see Fig. \ref{flow}), we need to find the
Voronoi partitioning of the lattice with point-to-point distance close to $a_{CC}=1.42{\text{~ \AA}}$. In principle, different type of interactions, other
than LJ, could be chosen for the pair interaction. However, LJ does a remarkably good job in describing the triangulation of the lattice
(as for any $fcc$-packed material, e.g. rare-gas solids at equilibrium).

The pseudospherical surface was truncated at a height corresponding to a radius equal to the lattice distance.
Particles were constrained to lie on a bounded surface. The combined use of a single-well potential ramping up immediately after a flat graphene ring surrounding the Beltrami pseudosphere and the bottom truncation circle eliminates the issue of the cluster interaction with the boundaries, thus eliminating the possibility for the latter to interfere with the internal formation of defects.

Due to an efficient implementation of these methods within the GSL numerical library, we first attempted a constrained minimisation using the conjugate gradient algorithm to find the ground-state triangular tiling of the pseudosphere. However, increasing the number of points results in several iterations to reach convergence, with often dramatic numerical instabilities and a rapidly unaffordable computational cost above a thousands points.

Thus, we decided to follow an alternative route and use the {\it Fast Inertial Relaxation Engine} (FIRE) approach [16] for finding the structures with minimum potential energy.
FIRE is a powerful global minimisation algorithm based on adaptive time step molecular dynamics, significantly more stable and faster than many sophisticated quasi-Newton schemes. In particular, structural relaxation is obtained by using the following equation of motion:
\begin{equation}\label{eqfire}
{\bf \dot v}(t)={\bf F}(t)/m -\gamma(t)|v(t)|({\bf \hat v}(t)- {\bf \hat F}(t))
\end{equation}
where $m$ is the mass (set fictitiously to 1 a.m.u. in our FIRE simulations), $v$ is the velocity and $F=-\nabla \phi$ is the force proportional to the gradient of the potential. In our case, $\phi$ is given by the LJ potential in Eq. \ref{LJ} and point-to-point distance is assessed according to the Beltrami's pseudosphere metric in Eq. 1. The adaptive time step spans the range from 0.001 to 1 fs, depending on the proximity to the energy minimum. Following Ref. 16 for all systems under investigation, the FIRE parameters have been set to the following: $N_{min}=5$, $f_{inc}=1.1$, $f_{dec}=0.5$, $\alpha_{start}=0.1$ and $f_{\alpha}=0.99$. Convergence below 1 meV/\AA~ was reached for all structures. For increasing number of atoms FIRE resulted at least three times faster than CG, while for the largest numbers (we reached up to 5 million lattice points by implementing FIRE on a GPU platform) was the only viable option due to memory and computational overload.

\section{Voronoi patterning and many-body classical potential}

Topological dualization of the generated structures provided models for $sp^2$-bonded carbon graphene sheets in the form of truncated pseudospherical surfaces.
The dualization process is performed by computing a triangulation of the surface. This was done by initially computing the adjacency matrix of each particle, where a neighbour was defined as a particle closer than $\sqrt{3}$ times the lattice distance. Distances were evaluated in three-dimensional space and not on the surface. The mesh was then refined in order to output a triangulation. The centres of each triangle were finally exported to a final structure containing pentagonal, hexagonal and heptagonal rings only.

\begin{figure}[h]
\centering
\includegraphics[width=0.5\linewidth]{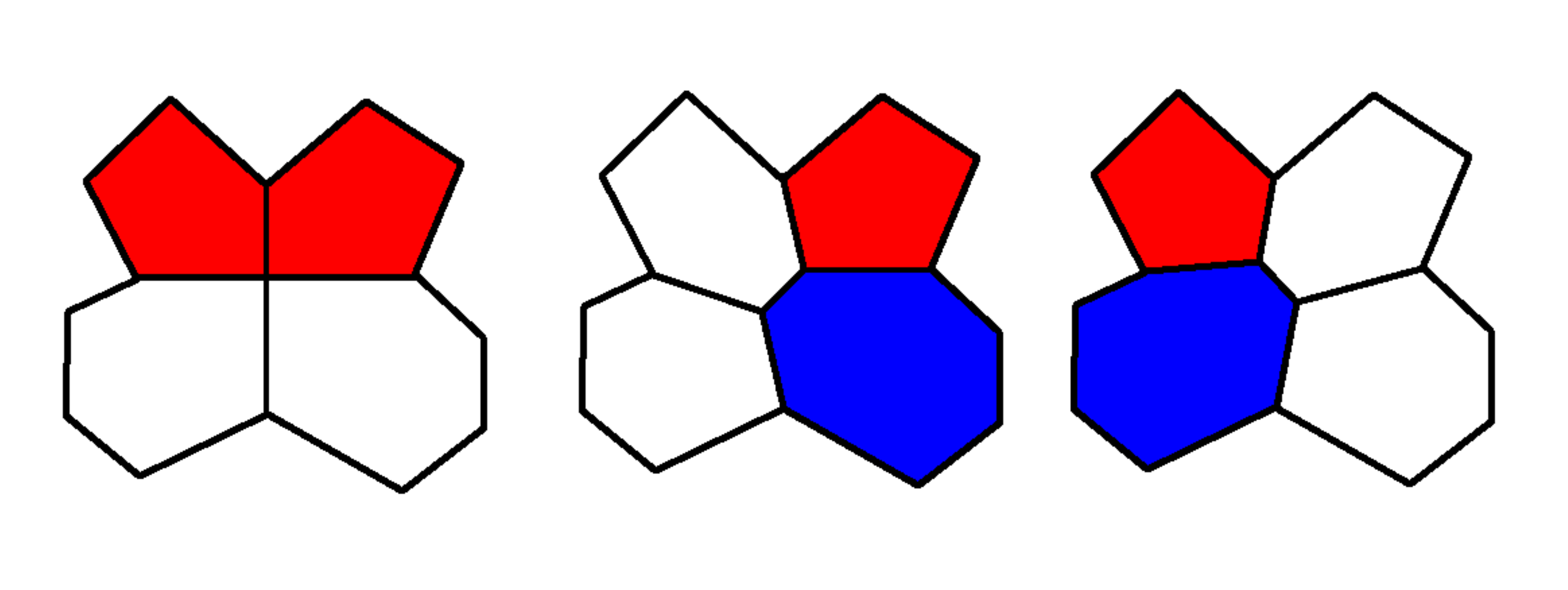}
\caption{A pentagon pair as shown in (a) is essentially an uncharged dipole. It that does not introduce topological curvature. An orthogonally oriented 5-7 pair resonating between two neighbour positions represents an alternative way to see this kind of defects (b and c).}
\label{pentpair}
\end{figure}

After the Voronoi patterning of the surface, the structure is not in its equilibrium position, and thus a re-optimization is necessary
by adding a three-body term to the model pair-wise LJ potential. In particular, this can be achieved
by expanding the many-body interaction energy $\phi_{MB}$ in a series of terms depending on atom pairs and atomic triplets as follows:
\begin{equation}\label{manybody}
\phi_{MB}=\frac{1}{2}\sum_i^N \sum_{j>i} \phi_{ij}(r_{ij})+\frac{1}{6}\sum_{i}^N \sum_{j>i}^N\sum_{k>j}^N\phi_{3B}(r_{ijk})
\end{equation}
for dealing with the covalent bonds connecting the carbon atoms. $\phi_{ij}(r_{ij})$ and $\phi_{3B}(r_{ijk})$ are the two-body and 3-body angular terms, respectively, and one-body terms that depend on external fields were suppressed in our simulations. In particular we adopted a Stillinger-Weber-type (SW) potential [16] with parameters optimised for carbon-based materials as follows: $A=5.373203$, $B=0.50824571$, $a=1.8943619$, $\lambda=18.707929$ and $\gamma=1.2$. The constant term appearing in the angular term is equal to 1/2 to favour ideal $sp^2$ configurations. To evaluate the efficiency of this potential in determining equilibrium configurations of carbon-based systems, we carried out the calculation of graphene cohesive energy.
This energy results equal to -7.459 eV, remarkably close to the value of -7.4 eV reported in literature from simulations using parametric interatomic potentials tailored for graphene.
The ground state of particles interacting via interatomic potentials on curved spaces displays an interesting set of features that originate from the interplay between local curvature and average inter-particle distance.
No vertex in these structures had a valence higher than 3. Vertex valence is an fundamental property to take into account when computing charge defect. Many metastable geometries were found during the minimization path. As an example, a defect of the type shown in Fig. \ref{pentpair} - although containing two pentagons - carries no net charge, and it would be actually equivalent to an uncharged 5-7 dipole. Surprisingly, other simulations also found the formation of pentagonal pairs.

Although we did not encounter vertices with a valence higher than 3, to avoid confusion, and in order to perform an exact computation of the topological charge involved, it might be useful to generalize the concept of number of sides of a polygon by including the valence of its vertices into it. We define the \emph{augmented} number of sides of a polygon as:

\begin{equation}
\tilde{n}_{i}=3\left(n_{i}-2\sum_{j=1}^{n_{i}}\frac{1}{v_{i}}\right)
\end{equation}

\begin{figure}[h]
\centering
\centerline{\includegraphics[width=.5\linewidth, trim=1cm 5cm 1cm 1cm]{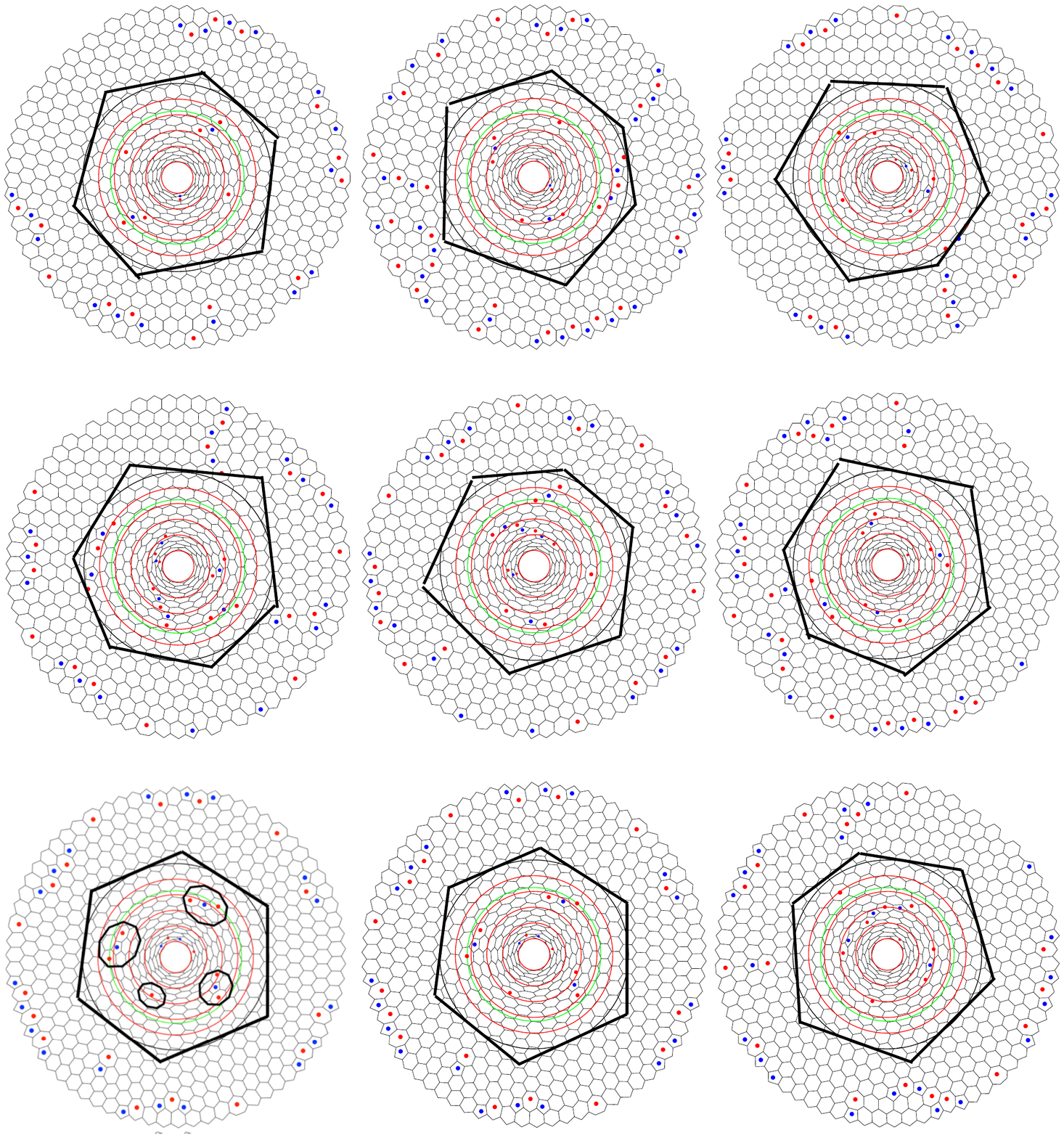}}
\caption{Beltrami pseudospheres with 1146 atoms obtained by dualization of a 614 point lattice with $R_{max}$=16 and $Z_{min}$=-23.}
\label{1146tot}
\end{figure}

\begin{figure}[hbt!]
\centering
\centerline{\includegraphics[width=.8\linewidth]{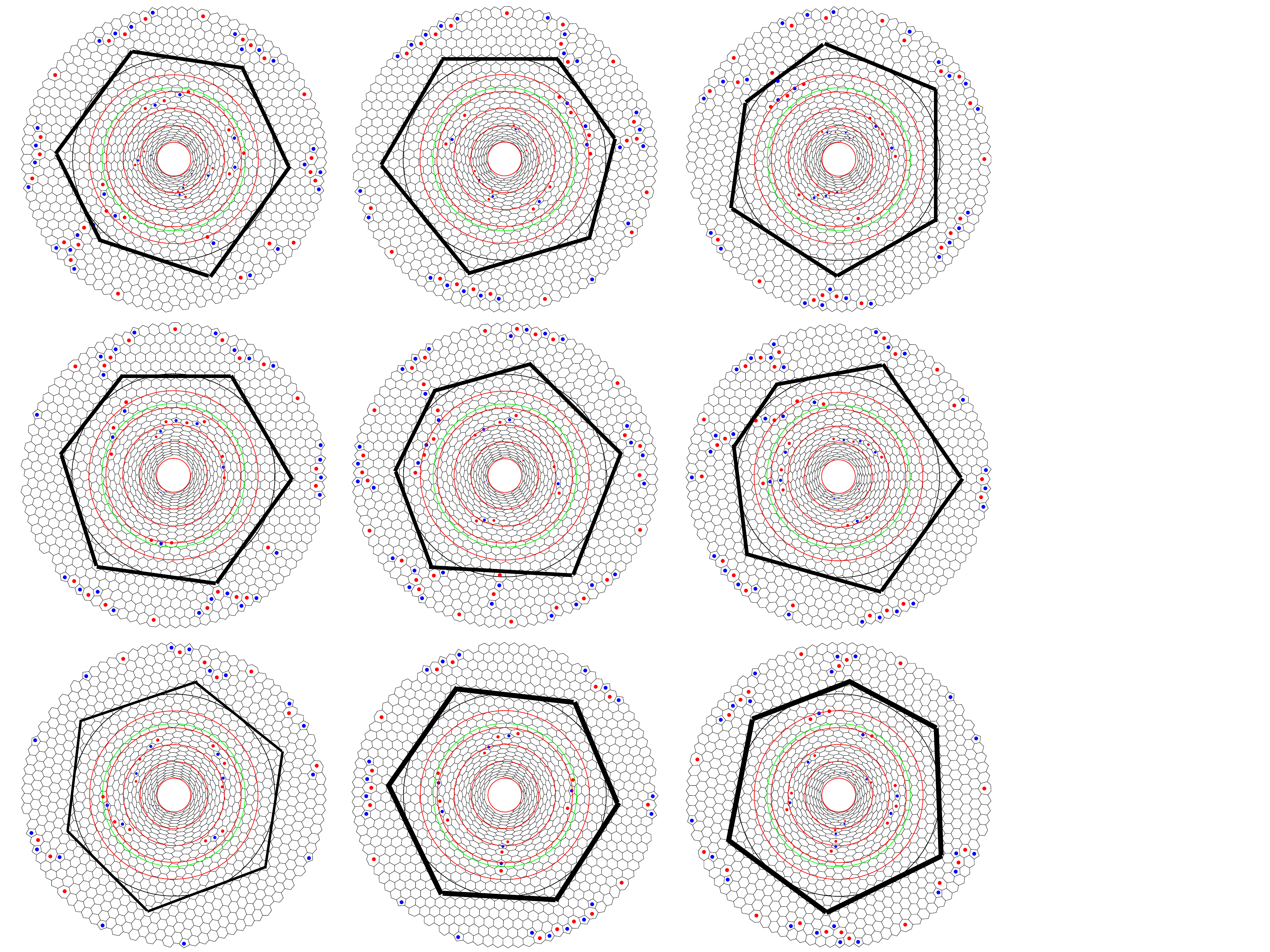}}
\caption{Beltrami pseudospheres with 2146 carbon atoms obtained by dualization of a 1128 point lattice with $R_{max}$=26 and $Z_{min}$=-38.}
\label{2146tot}
\end{figure}

\begin{figure}[hbt!]
\centering
\centerline{\includegraphics[width=.5\linewidth, trim=1cm 7cm 1cm 1cm]{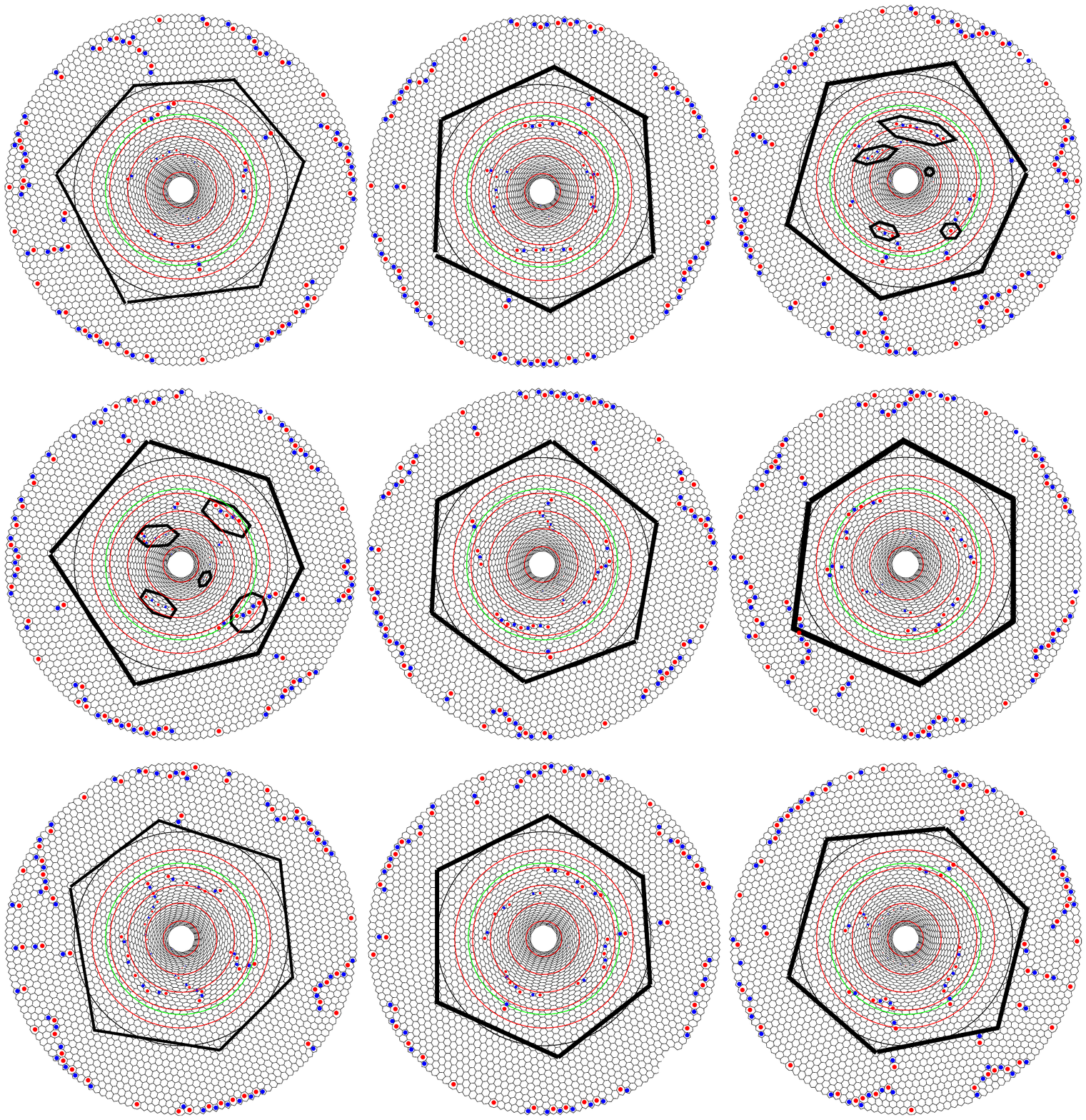}}
\caption{Beltrami pseudospheres with 5506 carbon atoms obtained by dualization of a 2840 point lattice with $R_{max}$=36 and $Z_{min}$=-64.}
\label{5506tot}
\end{figure}

Final optimized geometries for Beltrami pseudospheres of different size investigated in this work are reported in Figs. \ref{1146tot},\ref{2146tot},\ref{5506tot}. The Thomson
problem was solved for this set of constant negative curvature structures, as detailed in the text of the manuscript.

\section{Density Functional and Density Functional Tight-Binding}

To perform minimum energy calculations including explicitly the electronic motion into our simulations we used both the Density Functional and the Density Functional Tight-Binding approaches.

In particular, for the latter we adopted the self-consistent charge framework that leads to an improved description of the Coulomb interaction between atomic partial charges. Atomic interactions between carbon atoms were treated by the semirelativistic, self-consistent charge Slater-Koster parameter set matsci-0-3 [18]. To help convergence to the minimum energy, we employed a room temperature Fermi smearing of the electronic density. The unit cells in DFTB simulations were shaped as cubes with sides increasingly large for increasing number of atoms (70 \AA~ for 1146 carbon atoms, 90 \AA~ for 2146, and 130 \AA~ for 5506 carbon atoms, respectively). The advantage of using DFTB with respect to other {\it ab-initio} methods, such as Density Functional Theory (DFT), is due to the computational cost of this approach to electronic structure calculations, which is about two orders of magnitude cheaper than the corresponding full DFT calculation [17]. As a result of this substantial speed gain, DFTB may be used to investigate much larger systems than those accessible by DFT at an affordable computational cost.
Nevertheless, we checked the accuracy of DFTB against DFT in this case by performing structural optimisation with and without constraining the atoms to lay on the
Beltrami pseudosphere.
In the former case, fixed-nuclei energy difference calculations between two different Beltrami configurations, both consisting of 1146 carbon atoms, using DFT and DFTB were performed and found to be below 1 meV per carbon atom. DFTB can be then safely used in our simulations.

\begin{figure}[h]
\centering
\includegraphics[width=.75\linewidth,angle=-90]{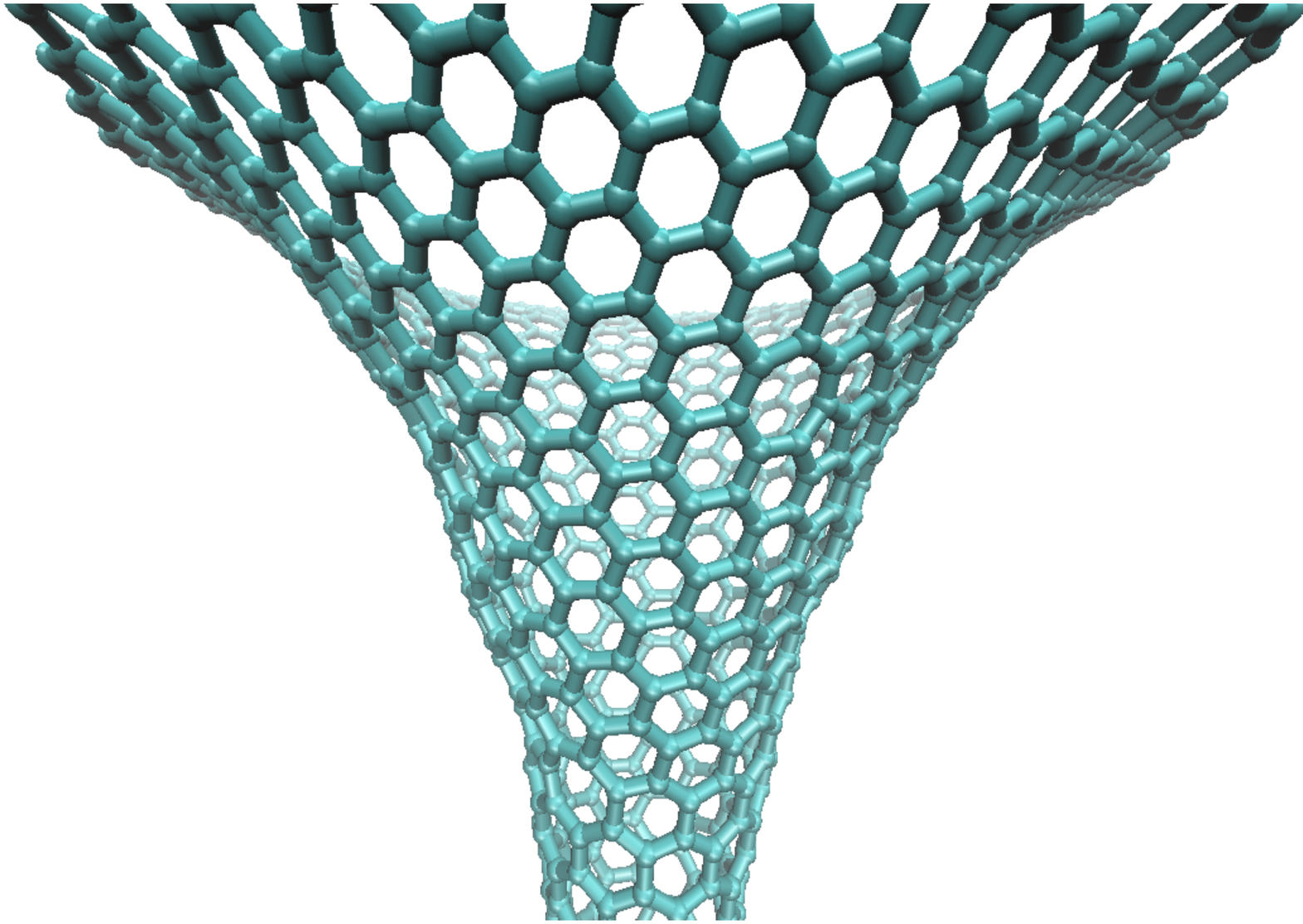}
\caption{Beltrami pseudospheres with 940 carbon atoms obtained by DFT optimization}
\label{DFT}
\end{figure}

Eliminating the bi-dimensional constraint, DFT was used to test the mechanical stability of these carbon-based structures. DFT calculations have been performed using the ab initio total-energy and molecular dynamics program VASP, with the implementation of an efficient extrapolation for the electronic charge density. The ion-electron interaction has been described using the projector augmented wave (PAW) technique, with single-particle orbitals expanded in plane waves with a cutoff of 400 eV, which ensures convergence
of the structural parameters of graphene, like the $a_{CC}$ distance to better than 0.05\% with respect to experimental data. We tested two different exchange-correlation functionals, notably LDA and PBE, and found no appreciable difference. Thus, after these tests, we decided to use the LDA functional. Thermal excitations of electrons were included via the finite-temperature formulation of DFT, in which the variational quantity to be minimized is the free energy of the electrons, $F_{static}=E-TS$, where the DFT energy $E$ is the usual sum of kinetic, electron-nucleus, Hartree, and exchange-correlation terms, and $S$ is the electronic entropy, given by the independent electron formula $S=-k_{\text B}T~\sum_i{[f_i \ln{f_i}+(1-f_i )\ln{1-f_i}]}$, with $k_{\text B}$ being the Boltzmann  constant and $f_i$ the thermal Fermi-Dirac occupation number of orbital $i$. In all calculations we used an electronic temperature of $T=300$ K. Brillouin zone sampling was performed using the $\Gamma$-point only. The electronic free energy of the unit cell, which was chosen as a cube of 55~\AA ~side, was converged to within less than 1 meV.
In Fig. \ref{DFT} we report the constraint-free structure obtained upon DFT optimization.
The intrinsic curvature induced locally by heptagonal defects is by far larger than what expected at any given location on the pseudosphere. For this reason, charged defects tend to spread over a wider area and often appear in the form of short segments, where the original curvature carried by a single heptagon is split into two halves and distributed at each end of the segment.

\end{document}